\documentclass[useAMS,usenatbib]{mnras}
\usepackage{graphicx,natbib,amssymb,amsmath,stmaryrd,makecell}
\usepackage{txfonts}
\usepackage{xcolor}
\usepackage{hyperref}
\usepackage{xspace}
\usepackage{caption}

\newcommand{\id}{{\,\rm d}}

\newcommand{\beq}{\begin{equation}}   %

\newcommand{\eeq}{\end{equation}}   %

\newcommand{\beqa}{\begin{eqnarray}}   %

\newcommand{\eeqa}{\end{eqnarray}}   %

\newcommand{\beal}{\begin{align}}

\newcommand{\enal}{\end{align}}

\newcommand{\bspl}{\begin{split}}

\newcommand{\espl}{\end{split}}

\newcommand{\bsub}{\begin{subequations}}

\newcommand{\esub}{\end{subequations}}

\newcommand{\bmulti}{\begin{multline}}   %

\newcommand{\beqm}{\begin{mathletters}}   %

\newcommand{\eeqm}{\end{mathletters}}   %

\newcommand{\Drr}{{\Delta \rho/\rho}}

\author[Acharya, Cyr and Chluba]
{Sandeep Kumar Acharya$^1$\thanks{E-mail:sandeep.acharya@manchester.ac.uk}, Bryce Cyr$^1$ and Jens Chluba$^1$
\\
$^1$Jodrell Bank Centre for Astrophysics, School of Physics and Astronomy, The University of Manchester, Manchester M13 9PL, U.K.
}
\date{\vspace{-0mm}Accepted XXX. Received YYY; in original form ZZZ}

\title[soft photon heating]
{The role of soft photon injection and heating in 21~cm cosmology}
\begin{document}

\maketitle

\begin{abstract}
    The ARCADE radio excess and EDGES measurement remain puzzling. A link between the two has been previously considered, however, in this work we highlight an important related effect that was not analyzed in detail before. By performing cosmological thermalization calculations with soft photon injection using {\tt CosmoTherm}, we show that for the 21~cm signal generation the interplay between enhanced radio spectral distortions and the associated heating can hide a significant radio excess before the reionzation era. We illustrate this effect for a simple power-law soft photon source in decaying particle scenarios. Even if simplistic, the uncovered link between CMB spectral distortions and 21~cm cosmology should apply to a much broader range of scenarios. This could significantly affect the constraints derived from existing and future 21~cm observations on the evolution of the ambient radio background. In particular, scenarios that would be ruled out by existing data without heating could become viable solutions once the heating is accounted for in the modelling. Our calculations furthermore highlight the importance of global 21~cm observations reaching into the dark ages, where various scenarios can potentially be distinguished.
\end{abstract}


%\section{CMB spectral distortion solutions as a function of injection frequency}
%---------------------------------------------------------------
\section{Introduction}

%---------------------------------------------------------------

The Cosmic Microwave Background (CMB) spectrum is well approximated by a Planck spectrum at a temperature $T_0=2.7255\,{\rm K}$, with possible distortions limited to $\Delta I/I \lesssim 10^{-5}-10^{-4}$ at frequencies $\nu \simeq 60-600\,{\rm GHz}$ \citep{Fixsen1996}. This measurement constrains exotic electromagnetic energy injection to the CMB such as dark matter decay or annihilation to standard model particles, direct photon injection etc. \citep{Sarkar1984, Hu1993b, Chluba2011therm,Bolliet2020PI}. In the pre-recombination ($z\gtrsim 10^3$) universe, addition of electromagnetic particles disturbs the photon-baryon equilibrium which result in broadband spectral distortion features such as $y$ or $\mu$-type distortions \citep{Zeldovich1969,Sunyaev1970mu}. At lower redshifts, photons and baryons decouple, and one expects inefficient thermalization. Since the measurements with {\it COBE/FIRAS} were limited to around the CMB maximum, there is an interesting possibility of injections of radio or soft photons at $\nu\lesssim 60$~GHz, which is not currently constrained by this measurement. Therefore, efforts towards new measurements of the CMB spectrum at increased frequency coverage and sensitivity have gathered a lot of attention with future experiment proposals such as {\it PIXIE} \citep{Kogut2011PIXIE, Kogut2016SPIE}, BISOU \citep{BISOU}, COSMO \citep{Masi2021}, TMS \citep{Jose2020TMS} or a spectrometer within the ESA Voyage 2050 space program \citep{Chluba2021Voyage}.  

 There are several complementary experiments which have measured the radio synchrotron background at $\simeq 10\,{\rm MHz}-10\,{\rm GHz}$ \citep{Fixsen2011,DT2018}. The measurements of both experiments were mostly consistent with each other, showing a detection of a radio excess with synchrotron-like power-law behaviour although at slightly different amplitudes. The intensity of this radio background is $\simeq 3-5$ times higher than the flux which can be accounted for by the CMB and discrete galactic and extragalactic sources \citep{GTZBS2008,Tompkins2022}, begging the question about what the origin of this signal may be. For a detailed discussions on potential galactic and extragalactic explanations of radio excess, the reader is referred to \citet{Singal2018,RSBworkshop2022}.
 
 While it is still a reasonable possibility that there are additional unaccounted radio sources, various authors have proposed new-physics solution for the radio excess \citep[see][for latest overview]{RSBworkshop2022}. Some of these proposed scenarios include Comptonized photon injection distortions \citep{Chluba2015GreensII, Bolliet2020PI}, annihilating axion-like dark matter \citep{Fraser2018}, dark photons \citep{PPRU2018, Caputo2022}, supernova explosions of population III stars \citep{JNB2019}, superconducting cosmic strings \citep{BCS2019}, the decay of relic neutrinos to sterile neutrinos \citep{CDFS2018}, thermal emission of quark nugget dark matter \citep{LZ2019}, bright luminous galaxies \citep{MF2019} and accreting astrophysical \citep{ECLDSM2018} or primordial black holes \citep{MK2021,ZYXMCL2023}. Some of these explanations may already be in tension with other cosmological data \citep{ADC2022,AC2022}, but still the model space is vast and additional data will be needed to reach a firm conclusion. 
 
 One consequence of the excess radio background on top of the CMB is that it gives rise to a stronger 21~cm absorption signal \citep[see][for a general review on 21~cm physics]{F2006} whose tantalizing signature at $z\simeq 20$ may have been reported by the EDGES collaboration \citep{Edges2018}. Although the EDGES observation is currently being debated \citep{HKMP2018,Saras2022}, and the hopes are high that ongoing and future experiments such as HERA \citep{HERA2022} and REACH \citep{REACH2022} will clarify the situation further, there is a possibility that both of these radio signals have a common physical origin. This was highlighted in \cite{FH2018} \citep[see also][]{FB2019} but the authors concluded that the radio background as seen today would be in tension with the EDGES measurement if it was simply redshifted back to $z\simeq 20$. Of course, the radio background could build up between $0< z\lesssim 20$ due to the presence of astrophysical sources which are not yet known. In the future, one may be able to study this option by measuring the upscattering of radio photons by hot electrons inside the galaxy clusters \citep{HC2021,LCH2022}; however, the current data does not shed much light on this possibility.

One of the key assumptions made in the analysis of \cite{FH2018} is that the radio photons just redshift and do not thermalize with the background electrons. This is not strictly true as some of the soft photons can be absorbed by the background electrons through the Bremsstrahlung/free-free process, thereby generating heat. The role of Bremsstrahlung in the thermalization of CMB spectral distortions at $z\gtrsim 2\times 10^6$ has been studied in \citet{Burigana1991,Hu1993,Chluba2015GreensII}. The efficiency of the absorption process is inversely proportional to the frequency of the soft photons. Therefore, for a synchrotron-like spectrum, most of the photons (in terms of energy) may still survive. However, the even lower frequency photons have a higher chance to be absorbed by the electrons. This heating will not create significant CMB distortion due to the weak coupling between photons and baryons at $z\lesssim 200$, but it can have potential consequences for the electrons since they are significantly outnumbered by the CMB photons themselves. Thus, indirectly this can have important consequences for the 21~cm signature, especially when scenarios with strong radio emission are being studied. The generation of heat due to photon absorption can furthermore raise the electron temperature to its ionization threshold, leading to collisional ionization and a reduced recombination rate, which modifies the CMB anisotropies that we see today. 

In this paper, we consider a toy-model of soft photon injection within a decaying-particle scenarios. We highlight that the EDGES signal and the observed radio background over $\simeq 10\,{\rm MHz}-10\,{\rm GHz}$ are no longer mutually-excluded when heating due to soft photon absorption is taken into account. We identify a reasonable parameter space for the considered model that is not excluded by existing data including the CMB constraints. A short description of the paper is given below. In Sec. \ref{sec:observables}, we briefly describe the cosmological observables that we consider in this work and in Sec. \ref{sec:implementation}, we provide some details regarding the implementation. We define our model parameters in Sect.~\ref{sec:model} and follow up with discussions of the constraints on these parameters in Sec.~\ref{sec:constraint}. We showcase the importance of soft photon heating in Sec.~\ref{sec:soft_heating} and conclude in Sec. \ref{sec:conclusion}.

%--------------------------------------------------------------
\section{Cosmological observables and constraints}
\label{sec:observables}
%-------------------------------------------------------------
In this section, we briefly review the cosmological observables that we use in this work to constrain the parameters of our phenomenological soft photon injection solutions.

%-------------------------------------------------------------
\subsection{CMB spectral distortions}
%------------------------------------------------------------
Spectral distortion of the CMB can be caused due to several processes, one example being electromagnetic energy injection in the pre-recombination epoch.  Energy injection into the CMB heats the electrons which in turn boosts the CMB photons creating a distortion on the CMB spectrum. The distorted spectrum is analytically given by a $y$-distortion \citep{Zeldovich1969} which is accurate in the low optical depth regime ($z\lesssim 10^4$). Repeated Compton scattering with the background electrons drives the CMB photons towards kinetic equilibrium, pushing the spectrum towards a Bose-Einstein distribution or $\mu$-distortion at $z\gtrsim 2\times 10^5$ \citep{Illarionov1975b}. Since Compton scattering is a photon-conserving process, energy injection to the CMB without a proportionate number of photons results in a $\mu$-distortion and not a Planck spectrum with the appropriate temperature. 

At $z\gtrsim 2\times 10^6$, photon non-conserving processes such as double Compton and Bremsstrahlung are very efficient, washing out any CMB distortions and thermalizing the spectrum to a Planckian \citep{Hu1993b,Chluba2011therm,Khatri2012b}. As we will see below, here we will not be interested in such high redshifts. Between the epoch of creation of a $y$-distortion ($z\approx 10^4$) and $\mu$-distortion ($z\approx 2\times 10^5$), there is an intermediate range where the CMB spectral distortions cannot be described as a linear combination of $y$ and $\mu$ and the distortion shape carries independent temporal information of the underlying energy injection process \citep{Chluba2011therm,Khatri2012mix}. To accurately capture the spectral evolution, we compute the spectral distortion solutions using {\tt CosmoTherm}, which follows all the important thermalization processes \citep{Chluba2011therm}. 

The current upper limit for the amplitude of the $y$ and $\mu$ parameters turns out to be $|y|\lesssim 1.5\times 10^{-5}$ and $|\mu|\lesssim 9\times 10^{-5}$, at $2\sigma$ \citep{Fixsen1996}. This translates to a constraint on the energy release as $\Drr\lesssim 6\times 10^{-5}$, where $\rho_{\gamma}$ is the CMB energy density today. This constraint depends upon the spectral shapes of distortions and can in principle change drastically for soft photon injections at $\nu\lesssim 60$ GHz in the post-recombination era. Here we will use the {\it COBE/FIRAS} data to place constraints on the distortion shapes from photon injection cases by computing the $\chi^2$ value with respect to the data. We deproject any temperature shift from the obtained distortion signal.

%------------------------------------------------------------
\subsection{CMB anisotropy}
%------------------------------------------------------------
The observed CMB temperature and polarization anisotropies sensitively depend upon the recombination history and thermal properties of the universe. Over the decades, our knowledge of the standard recombination history has improved \citep{Zeldovich68,Peebles68,Seager2000,Sunyaev2009,chluba2010b,Yacine2010c} and can now be computed extremely precisely. Injection of energetic electrons/positrons and photons around the recombination epoch can heat and ionize neutral hydrogen and helium, modifying the recombination history. This increases the freeze-out fraction of free electrons at redshift $z\lesssim 10^3$. The increased probability for scattering of the CMB photons with the free electrons damps the CMB temperature anisotropy, while boosting the polarization signals \citep{ASS1999,Chen2004}. Precise observation of the CMB anisotropies \citep{WMAP7yrPower,Planck2018params} thus place strong constraints on these type of energy injection cases \citep{Galli2009,Slatyer2009,Huetsi2011,AK2019}. 

We compute the recombination history including these energy injection cases using the {\tt Recfast++} \citep{chluba2010b} module within {\tt CosmoTherm} \citep{Chluba2011therm}. One important addition is that we carefully take into account the heating/cooling caused by soft photons that are absorbed by the free-free process at low frequencies, $x\ll 1$. This leads to an important direct interplay between soft photon sources and the heating of the medium \citep{Chluba2015GreensII, Bolliet2020PI}, which we explore here for 21~cm signals.

%----------------------------------------------------------
\subsection{21~cm distortion}
%----------------------------------------------------------
The 21~cm brightness distortion measured against the background at redshift $z$, is given by its differential brightness temperature \citep[see e.g.][]{F2006},
%----------------------------------------
\begin{equation}
\label{eq:DTb}
    \Delta T_\text{b} = \dfrac{\left(1-{\rm e}^{-\tau_{21}}\right)}{1+z}\left(T_\text{s}-T_\text{R} \right)
   % \label{eq:21~cm_distortion}
\end{equation}
%----------------------------------------
where $T_\text{R}$ is the radiation temperature at the 21~cm resonance, $T_\text{s}$ is the spin temperature and $\tau_{21}$ is the 21~cm optical depth. In standard cosmology, it is usually assumed that the CMB is the only source of radio photons, i.e., $T_{\rm R}=T_{\rm CMB}$. However, with additional radio photons, we need to use the radiation temperature $T_{\rm R}=T_{\rm CMB}+\Delta T$. In the Rayleigh-Jeans regime, we can write,
%----------------------------------------
\begin{equation}
    \frac{T_{\rm CMB}+\Delta T}{T_{\rm CMB}}\Bigg|=\frac{I_{\rm CMB}+\Delta I}{I_{\rm CMB}}\Bigg|_{1.4 {\rm GHz}},
\end{equation}
%----------------------------------------
where $\Delta I/I_{\rm CMB}|_{1.4 {\rm GHz}}$ is the CMB spectral distortion in the rest frame frequency of 1.4 GHz at $z$. 

The signal is observed in absorption when $\Delta T_\text{b} <0$ and emission when $\Delta T_\text{b} >0$. The spin temperature in turn is defined by the ratio of the population of the upper and lower hyperfine states:
%----------------------------------------
\begin{equation}
n_1/n_0 \equiv 3 {\rm e}^{-T_\star/T_\text{s}}
\end{equation}
%----------------------------------------
where $T_\star = h\nu_{21}/k_\text{B}=0.068\text{~K}$, $\nu_{21} = 1.42 \text{ GHz}$, $h$ is the Planck constant, $k_\text{B}$ is the Boltzmann constant and the constant $3$ is a statistical degeneracy factor. The spin temperature is given by \citep{Venu2018}
\begin{equation}
T_\text{s} ^{-1}= \dfrac{x_\text{R} T_\text{R}^{-1} + x_{\rm c} T_\text{M}^{-1} + x_\alpha T_\alpha^{-1}}{x_\text{R}+ x_{\rm c} +x_\alpha},
\label{eq:spin_temperature}
\end{equation}
where $x_{\rm R}$, $x_{\rm c}$ and $x_\alpha$ are the radiative, collisional and WF coupling coefficients respectively, $T_\text{M}$ is the matter temperature, and $T_\alpha$ is the colour temperature of the Ly$\alpha$ radiation field. 

The EDGES collaboration \citep{Edges2018} has claimed detection of a 21 cm absorption feature with $\Delta T_{\rm b}\simeq-500$~mK originating from $z\approx 18$ and a 1$\sigma$ error of 200 mK. 
In this paper, we demand that $-500~{\rm mK}\lesssim \Delta T_{\rm b} \lesssim 0$ at $z\approx 18$. Outside of this regime we use a Gaussian likelihood with an error of 200 mK to quantify the tension with this data. For our calculations, this constraint usually only kicks in at $\Delta T_{\rm b}\lesssim -500$~mK, as none of our models have significant $\Delta T_{\rm b}>0$ at $z=18$. 
We again remind the reader that these are not robust constraints and we use them as a figure of merit.

\vspace{-3mm}
%---------------------------------------------------------
\subsection{Radio synchrotron background (RSB) excess}
%---------------------------------------------------------
The ARCADE-2 experiment \citep{Fixsen2011} measured a RSB between 3-90 GHz. These RSB data points alongside some previous results which were analyzed and compiled in their Table 4, are well fit by a power law with spectral index 2.6 and temperature $T\simeq 24$~K at 310 MHz. In \cite{DT2018}, the authors redid this analysis using independent data points around $\simeq 40-80$ MHz and found the best fit slope to be consistent with ARCADE but with slightly higher normalization of $\simeq$ 30K at 310 MHz. In this paper, we use the data points compiled in Table 2 of \cite{DT2018}.

In the aforementioned analyses, the contribution from resolved extra-galactic sources were not taken into account. However, in this work, we use a fitting function to the minimal extra-galactic background (MEG) which is given by \citep{GTZBS2008},
%---------------------------------------------------------
\begin{equation}
    T_{\rm bg}(\nu)\simeq 0.23\,{\rm K}\left(\frac{\nu}{\rm GHz}\right)^{-2.7}. 
    \label{eq:extra-galactic}
\end{equation}
%---------------------------------------------------------
This background is added to the injected photon spectrum to obtain a fit to the RSB data. We obtain a power-law fit with 
%---------------------------------------------------------
\begin{equation}
\label{eq:RSB-fit}
    T_{\rm RSB}(\nu)\simeq 1.230\,{\rm K}\left(\frac{\nu}{\rm GHz}\right)^{-2.555}. 
\end{equation}
%---------------------------------------------------------
when combining both data sets. We use this model as a reference for our likelihood evaluation. 

We note that recently it was pointed out that when including the minimal background one expects a power-law model with curvature to provide a better fit to the RSB data \citep{AC2022}. Additional data points and more precise data at $\nu\gtrsim$ 1 GHz may furthermore help distinguish between different explanations. By comparing the distortion obtained with {\tt CosmoTherm} to the RSB we then search for viable solutions.

%-----------------------------------------------------
\section{Implementation of the problem}
\label{sec:implementation}
%----------------------------------------------------

Here, we discuss some of the technical aspects of the calculations briefly, but refer to \citep{ADC2022} for details. 

%---------------------------------------------------
\subsection{21 cm modelling}
%---------------------------------------------------

Our modelling of the spin temperature ($T_S$) is carried out in a fashion similar to \cite{Mittal2022}. The main effects that we include are interactions of neutral hydrogen with the radio background through radiative coupling, collisions with hydrogen atoms, electrons and protons and Wouthuysen-Field coupling. We solve the  radiative coupling coefficient iteratively similar to \cite{Mittal2022}. Our modelling of the Lyman-alpha background follows from \cite{Hirata2006}. We assume only stellar contributions to the Lyman-alpha background with comoving emissivity given by,
%--------------------------------------------------
\begin{equation}
    \epsilon_\alpha(E,z) = f_\alpha \phi_\alpha(E) \dfrac{\dot{\rho}_\star(z)}{m_\text{b}}
\end{equation}
%-------------------------------------------------
where $f_\alpha$ is a scaling factor for the strength of the Ly$\alpha$ background, $\dot{\rho}_\star$ is the comoving star formation rate density (SFRD) and $m_\text{b}$ is the number-averaged baryon
mass. The collisional coupling coefficient $x_{\rm c}$ can be calculated using the recombination history from {\tt Recfast++} and spin-exchange rate coefficients tabulated in literature which can be found in \cite{ADC2022}.

%---------------------------------------------------
\subsection{Reionization modelling}
%---------------------------------------------------

We model reionization following the treatment of \cite{F2006}. The evolution of the hydrogen and helium fractions is given by,
%-------------------------------------------------
%-------------------------------
\begin{subequations}
\begin{align}
    \dfrac{\id x_{\rm HII}}{\id t} &= \xi_\text{ion}(z)\dfrac{\id f_\text{coll}}{\id t} - \alpha_A C\,x_{\rm HII} \,n_{\rm e} 
    \\
    \dfrac{\id x_{\rm HeII}}{\id t} &= \xi_\text{ion}(z)
    \dfrac{\id f_\text{coll}}{\id t} - \alpha_A C\,x_{\rm HeII} \,n_{\rm e},
\end{align}
\end{subequations}
%-------------------------------
where $x_{\rm HII}=\frac{n_{\rm HII}}{n_{\rm H}}$, $x_{\rm HeII}=\frac{n_{\rm HeII}}{n_{\rm H}}$ are ionized fractions, $f_\text{He}=\frac{n_{\rm He}}{n_{\rm H}}$ is the fraction of helium, $\xi_\text{ion}$ is the ionizing efficiency parameter, $f_\text{coll}$ is the matter collapse fraction, $\alpha_A$ is the case-A recombination coefficient, $C \equiv \langle n_{\rm e}^2\rangle/\langle n_{\rm e}\rangle^2$ is the clumping factor, and $n_{\rm e}$ is the total electron number density. We use the fitting function of \cite{SHTS2012} for the clumping factor.

The ionizing efficiency parameter is given by
%-------------------------------
\begin{equation}
    \xi_\text{ion} = A_\text{He} f_\star f_\text{esc}N_\text{ion}
\end{equation}
%-------------------------------
where $A_\text{He}$ is a correction factor due to the presence of helium, $N_\text{ion}$ is the number of ionizing photons per baryon, $f_\text{esc}$ is the fraction of ionizing photons escaping the host halo and $f_\star$ is the star formation efficiency. The expression for the X-ray heating of electrons is given by \citet{F2006}, 
\begin{equation}
    \dfrac{2}{3}\dfrac{\epsilon_X}{k_\text{B}n_{\rm H} H(z)} = 10^3\text{K}~f_X \left[\dfrac{f_\star}{0.1}\right]
    \left[\dfrac{f_{X,h}}{0.2}\right]
    \left[\dfrac{\id f_\text{coll}/\id z}{0.01}\right]
    \left[\dfrac{1+z}{10}\right]
\end{equation}
where $f_{X,h} \simeq (1+2x_{\rm e})/3$ is the fraction of X-ray energy used in heating \citep{Chen2004}, and $f_X$ is a scaling factor. 

To produce a reionization history where the universe is completely ionized at $z\simeq 6-7$, we use the following combination of parameters ($N_\text{ion}$, $f_\star$, $f_\text{esc}$, $f_\alpha$, $f_X$) = (4000, 0.1, 0.1, 1.0, 1.0). However, from the expression for $\xi_\text{ion}$, it is clear that this combination is not unique, and other choices may reproduce the same reionization history.

%---------------------------------------------------
\subsection{Principal Component Analysis for CMB anisotropy}
%---------------------------------------------------
To compute CMB anisotropy constraints, we use the direct projection method developed in \cite{PCA2020}. This method is based on a principal component analysis (PCA) of recombination history perturbations first developed in \citet{Farhang2011, Farhang2013} but used here with updated data from {\it Planck} 2018 \citep{Planck2018params}. With {\tt CosmoTherm} we can compute the changes to the standard ionization history of the universe, $\xi(z)=\Delta x_{\rm e}/x_{\rm e}$, for any given injection history. We can then compute the first three principal component coefficients by projecting $\xi(z)$ onto the eigenmodes, $E_i(z)$, with the integral
%--------------------------------------------------
\begin{equation}
\mu_i=\int \xi(x)E_i(z) \,{\rm d}z.
\end{equation}
%-------------------------------------------------
Using the covariance matrix of the $\mu_i$ obtained in \cite{PCA2020} we can then compute the likelihood of the model assuming Gaussian statistics. This provides a fast way of estimating how consistent the obtained ionization history is with the CMB anisotropy data of {\it Planck}, mimicking the results from a full MCMC analysis.

%------------------------------------------------------------
\begin{figure}
\centering 
\includegraphics[width=\columnwidth]{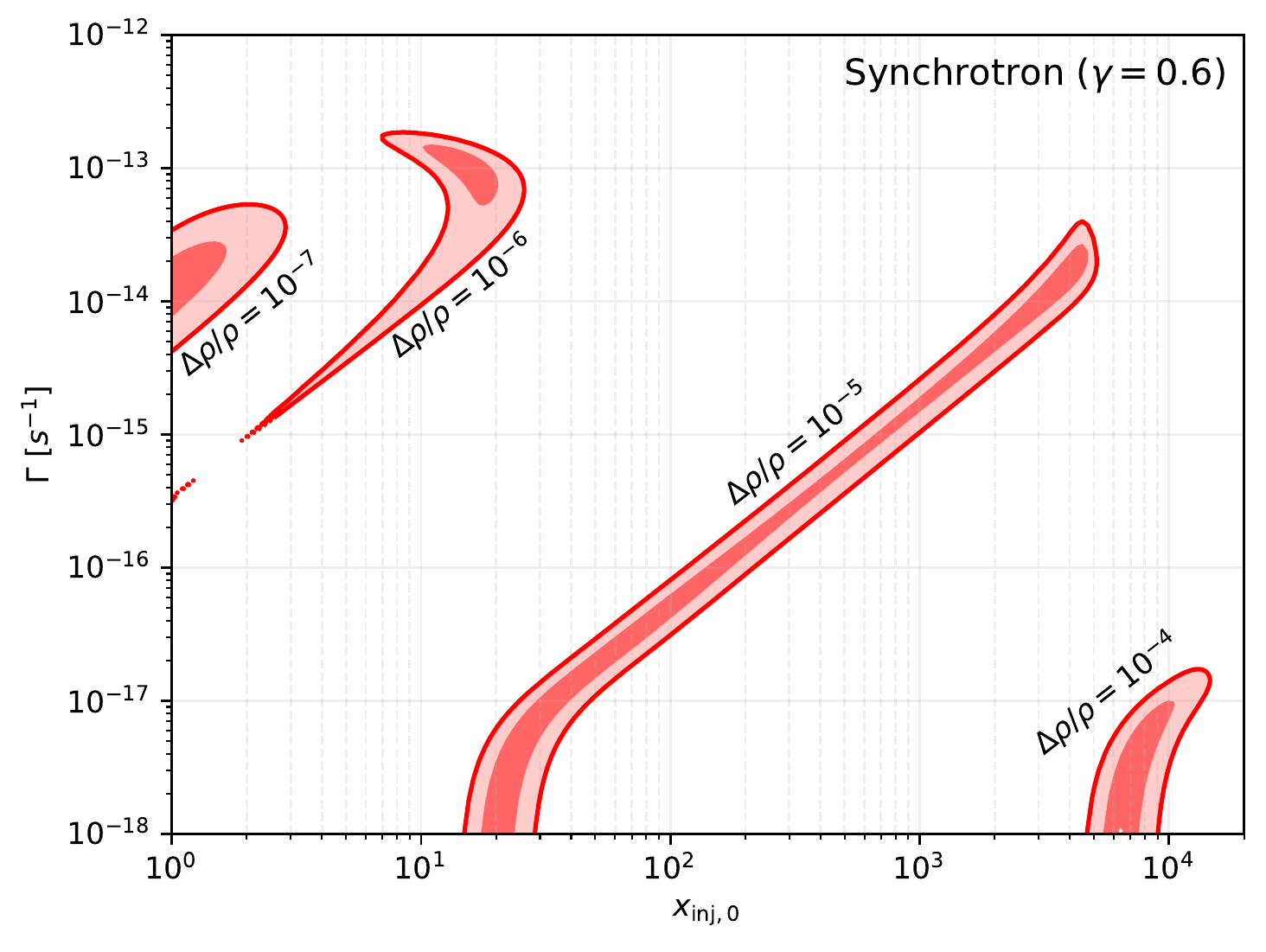}
\hspace{0mm}
%\includegraphics[width=\columnwidth]{./eps/Synchrotron_constr.pdf}
%\\[1mm]
\caption{Constraints on the parameter space of synchroton-like injection problems for a few different values of $\Drr$. This is the joint constraint using CMB and 21~cm distortions, as well as observed RSB data. Models within the dark red contours are found to be consistent within $1\sigma$ to the best fit model of the data, while light red contours are within $2\sigma$. }
\label{fig:sync_constraint}
\end{figure}
%------------------------------------------------------------

The eigenmodes $E_i(z)$ are sensitive to changes in the ionization history at high redshifts $100\lesssim z \lesssim 4\times 10^3$ \citep{PCA2020} but do not capture changes to reionization history at lower redshifts. Consideration of modifications to the reionization history are expected to further tighten the bounds; however, the uncertainties in the reionization model are significant and we do not use this information here.
Even if the details might change, the main conclusion should not be affected by this.

%--------------------------------------------------
\section{Phenomonological model for soft photon injection}
\label{sec:model}
%-------------------------------------------------
In this work, we consider energy injection in the form of soft photons envisioning the decay of a light dark matter candidate. We have chosen this example to simplify the calculations and give a flavour of the basic physics relating to soft photon heating. In this approach, the time dependence of the injection process is fully defined by the decay rate. The energy injection rate to the CMB is given by,
%--------------------------------------------------
\begin{equation}
    \frac{{\rm d}\rho_{\gamma}}{{\rm d}t}=f_{\rm dm}\Gamma\rho_{\rm cdm}{\rm e}^{-\Gamma t},
    \label{eq:decay}
\end{equation}
%--------------------------------------------------
where $f_{\rm dm}$ is the fraction of decaying dark matter, $\rho_{\rm cdm}$ is the energy density of cold dark matter and $\Gamma$ is the inverse of decay lifetime. 

For monochromatic photon injections, constraints on $f_{\rm dm}$ from spectral distortions were obtained in \cite{Bolliet2020PI}. Here instead we consider the case where the decay products exhibit an extended soft photon spectrum with a high frequency cutoff.  We assume the soft photon spectrum to be of the form,
%---------------------------------------------------
\begin{equation}
    \Delta I(x)\propto x^{-\gamma}{\rm e}^{-\frac{x}{x_{\rm inj,0}}},
    \label{eq:soft_spectrum}
\end{equation}
%---------------------------------------------------
where $x=h\nu/k_{\rm B}T_{\rm CMB}(z)$, $x_{\rm inj,0}=E_{\rm inj}/k_{\rm B}T_{\rm CMB,0}$ with $T_{\rm CMB,0}$ being the CMB temperature today and $E_{\rm inj}$ a cutoff energy that is related to the energetics of the photon production process. The redshifting of the photons is implicitly taken into account in the definition of the dimensionless frequency $x$. Therefore, a photon emitted at $x$ and any redshift will show up today at the same $x$ if we ignore absorption or Comptonization. We clarify that $x$ is a variable while $x_{\rm inj,0}$ and subsequently $E_{\rm inj}$ are parameters in our model. The amplitude of the soft photon spectrum is determined by the combination of $\Delta \rho/ \rho_{\gamma}$, $x_{\rm inj,0}$ and $\Gamma$. We consider a free-free and synchrotron spectrum as example cases with $\gamma=0$ and 0.6, respectively. 

Even if we are choosing a decay-like time-dependence for the injection mechanism, we note that typical astrophysical emission processes exhibit a power-law spectrum \citep{Blumenthal1970, Rybicki1979}. One also expects free-free type emission from high energy particle cascades, in particular in dense environments, where a lot of reprocessing of the injected energy occurs. Our choice of spectrum therefore captures a broad range of physical scenarios and we plan to study constraints on the fraction of soft to hard photons in a forthcoming paper (Acharya et al., 2023).

%------------------------------------------------------------------
\section{Constraint on the model parameter space}
\label{sec:constraint}
%------------------------------------------------------------------

%------------------------------------------------------------
\begin{figure*}
\centering 
\includegraphics[width=\columnwidth]{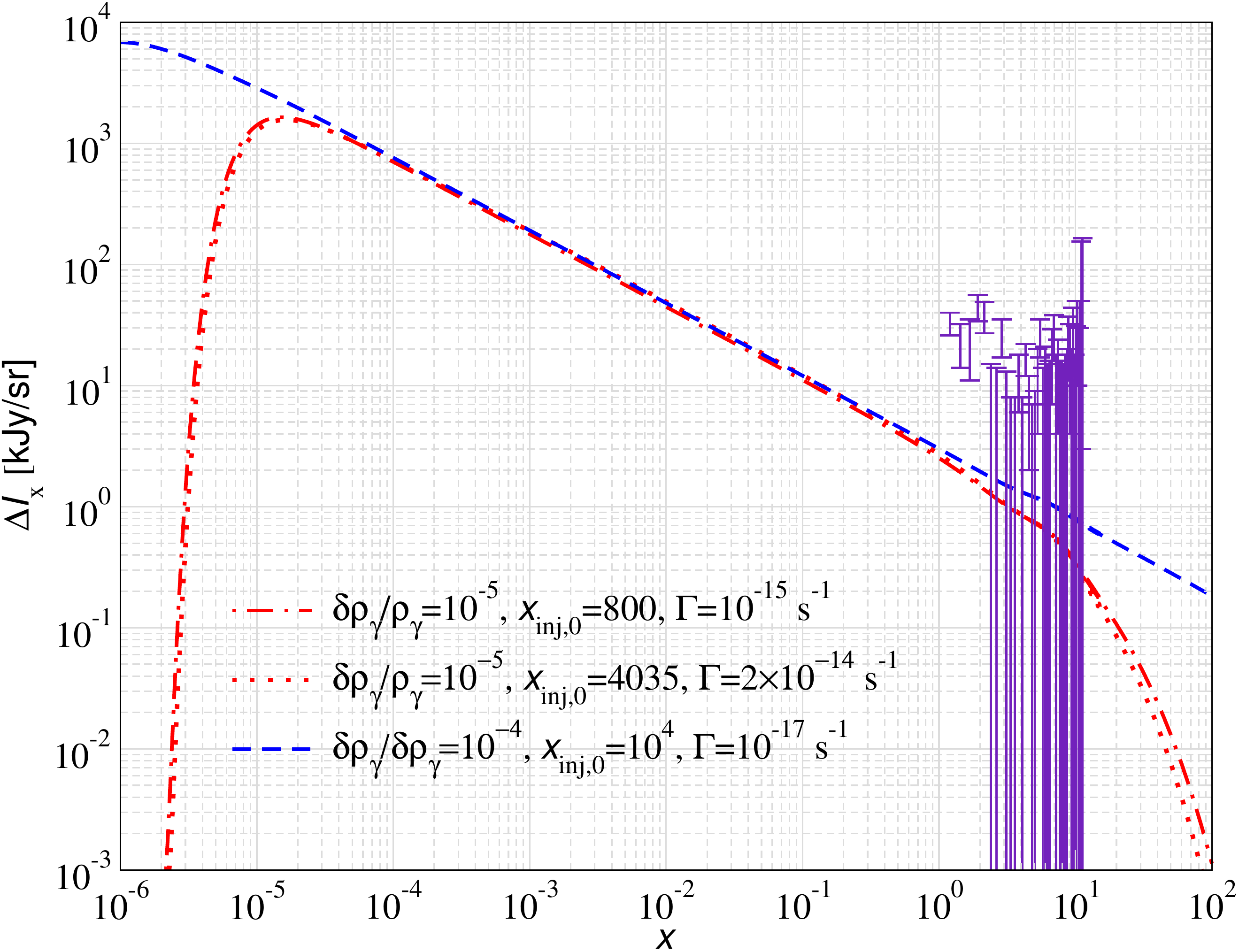}
\hspace{4mm}
\includegraphics[width=\columnwidth]{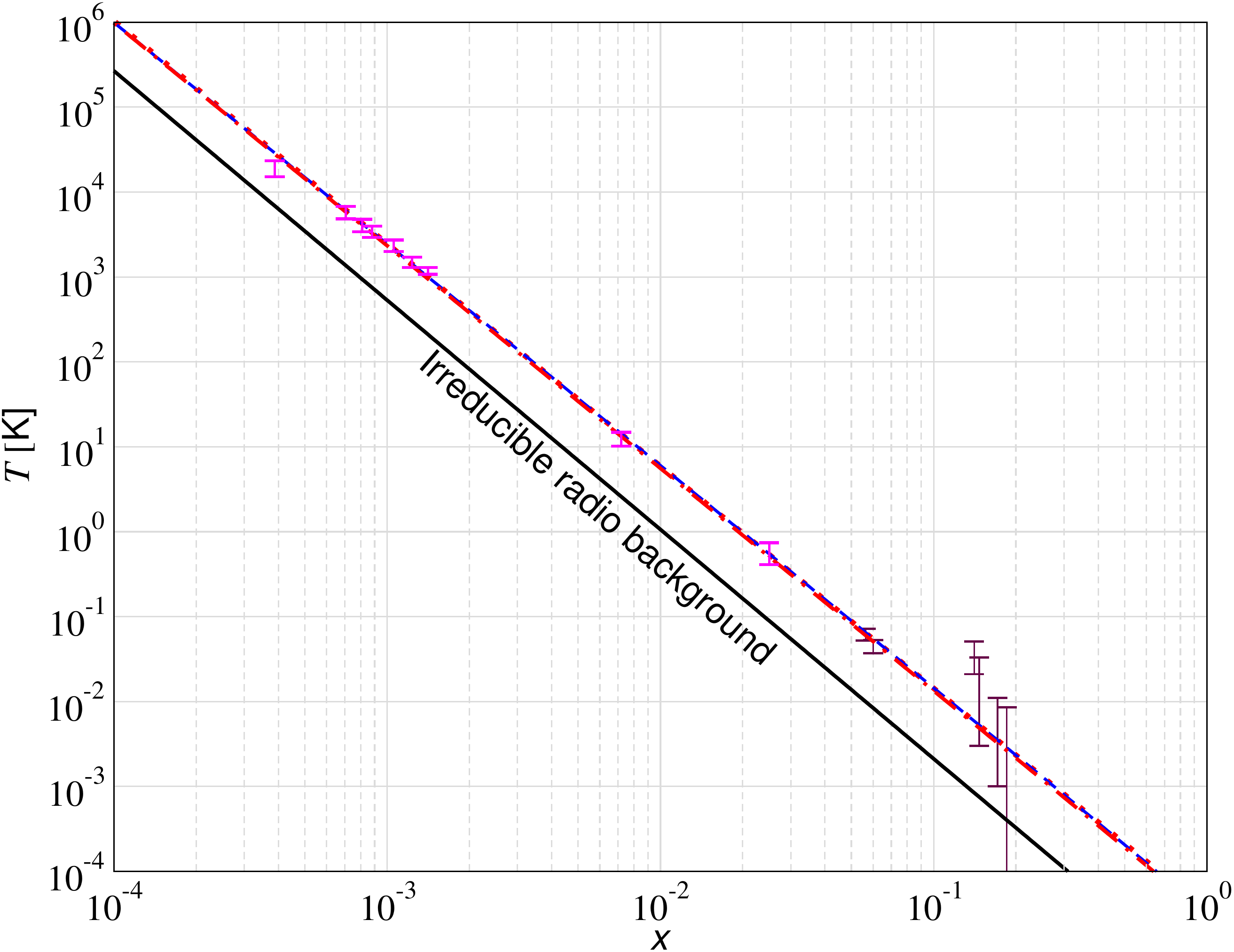}
\\[10mm]
\includegraphics[width=\columnwidth]{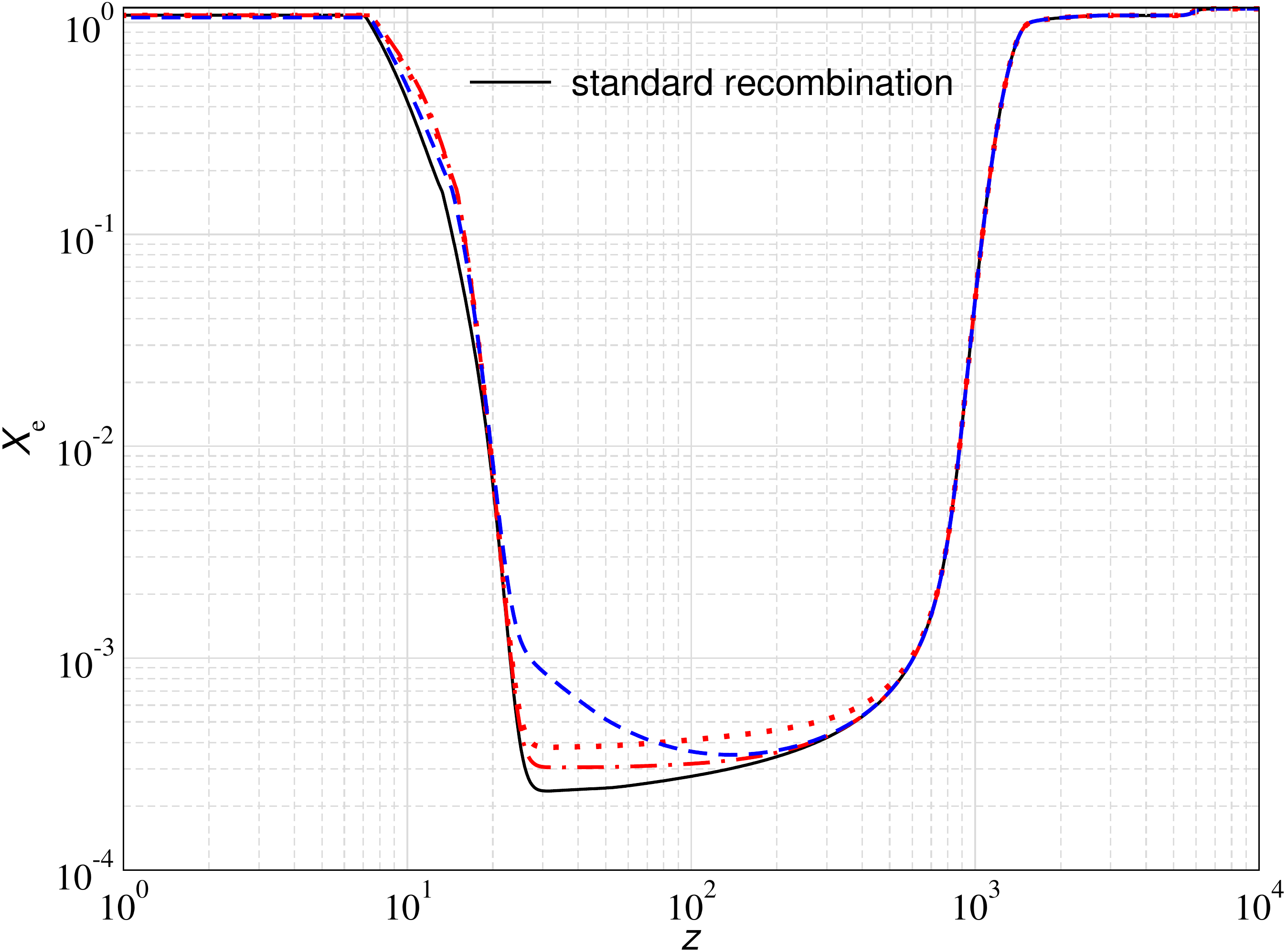}
\hspace{4mm}
\includegraphics[width=\columnwidth]{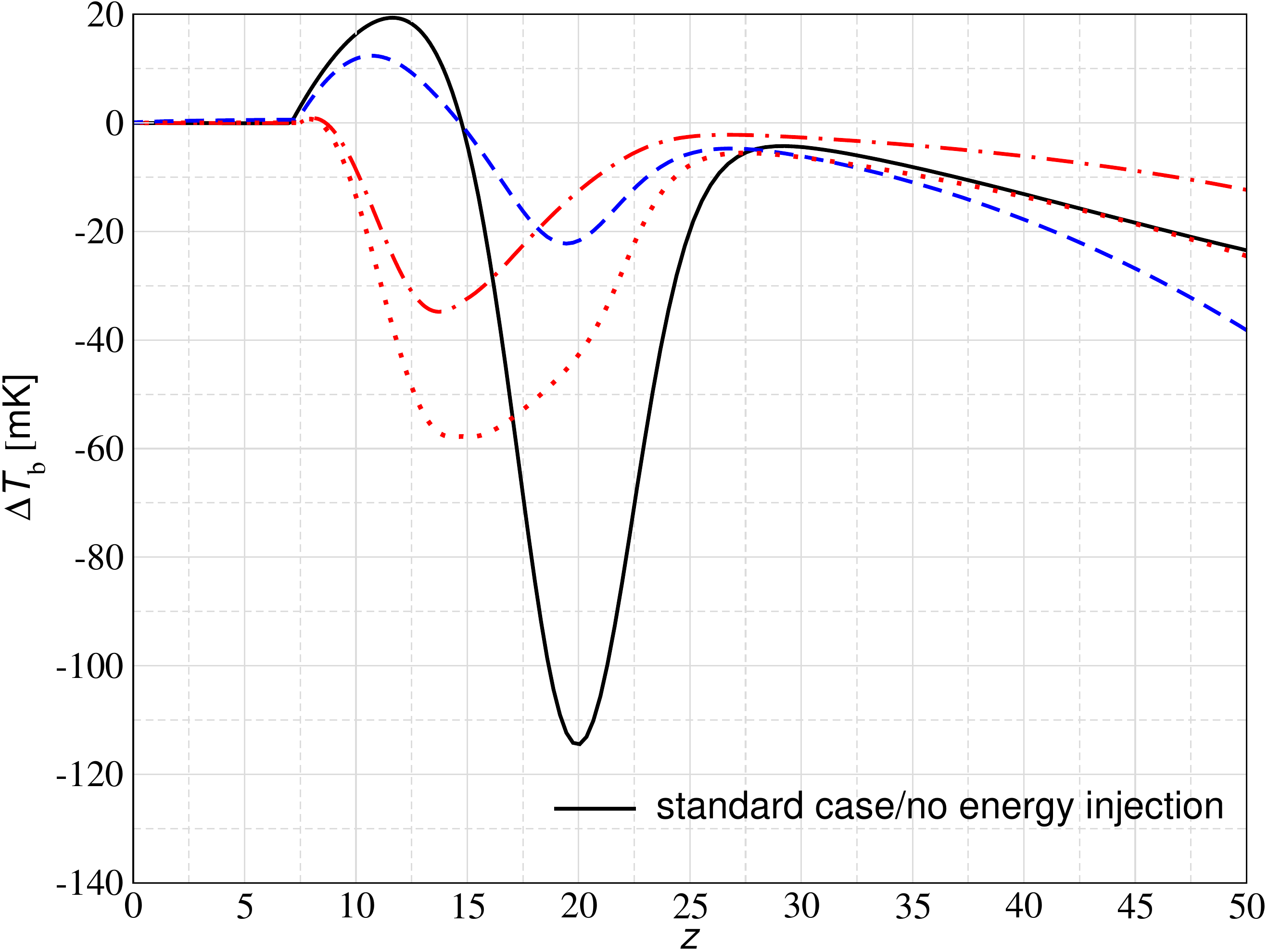}
%\\[10mm]
\caption{Dimensionless intensity (upper left), equivalent temperature in the Rayleigh-Jeans limit (upper right), ionization history (bottom left) and  the 21 cm distortion (bottom right) for a few parameter combinations which fit the data well for a synchrotron-like spectrum with parameter combinations as shown. We show the CMB spectral data \citep{Fixsen1996} in violet and for the RSB data, \citep{DT2018} is in magenta while ARCADE-2 \citep{Fixsen2011} is in maroon bands.}
\label{fig:soln_fit}
\end{figure*}
%------------------------------------------------------------

%------------------------------------------------------------
\begin{figure*}
\centering 
\includegraphics[width=\columnwidth]{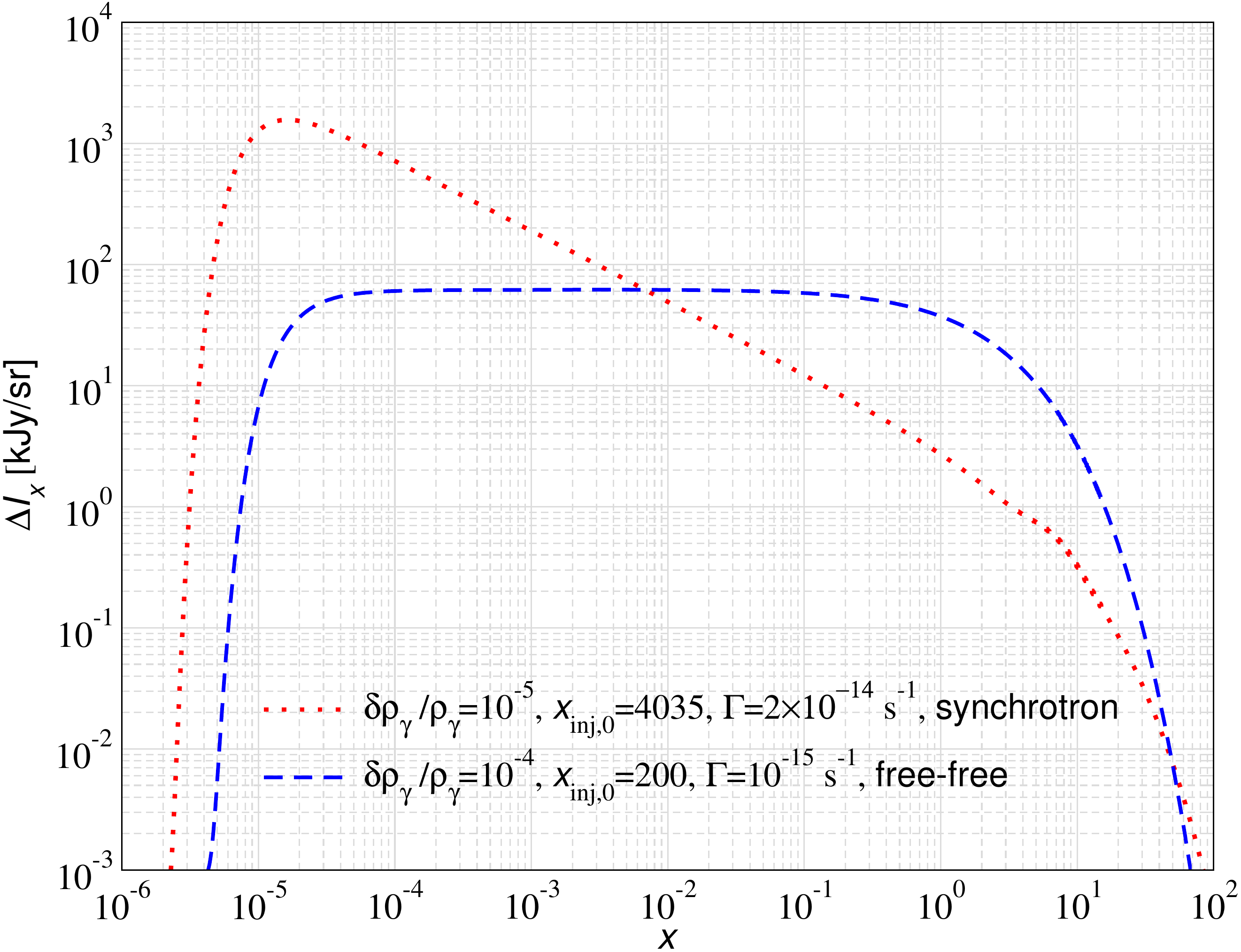}
\hspace{4mm}
\includegraphics[width=\columnwidth]{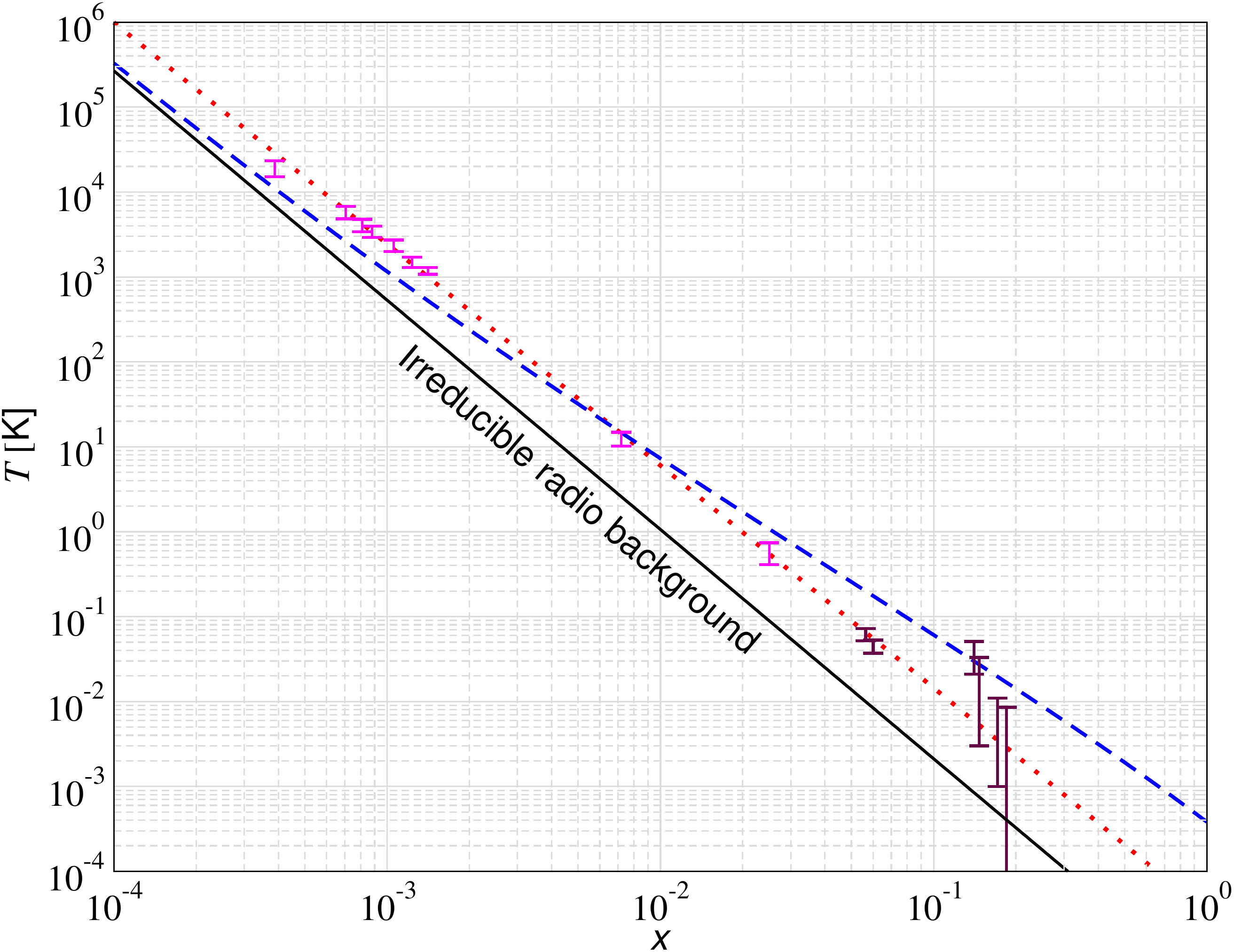}
\\[10mm]
\includegraphics[width=\columnwidth]{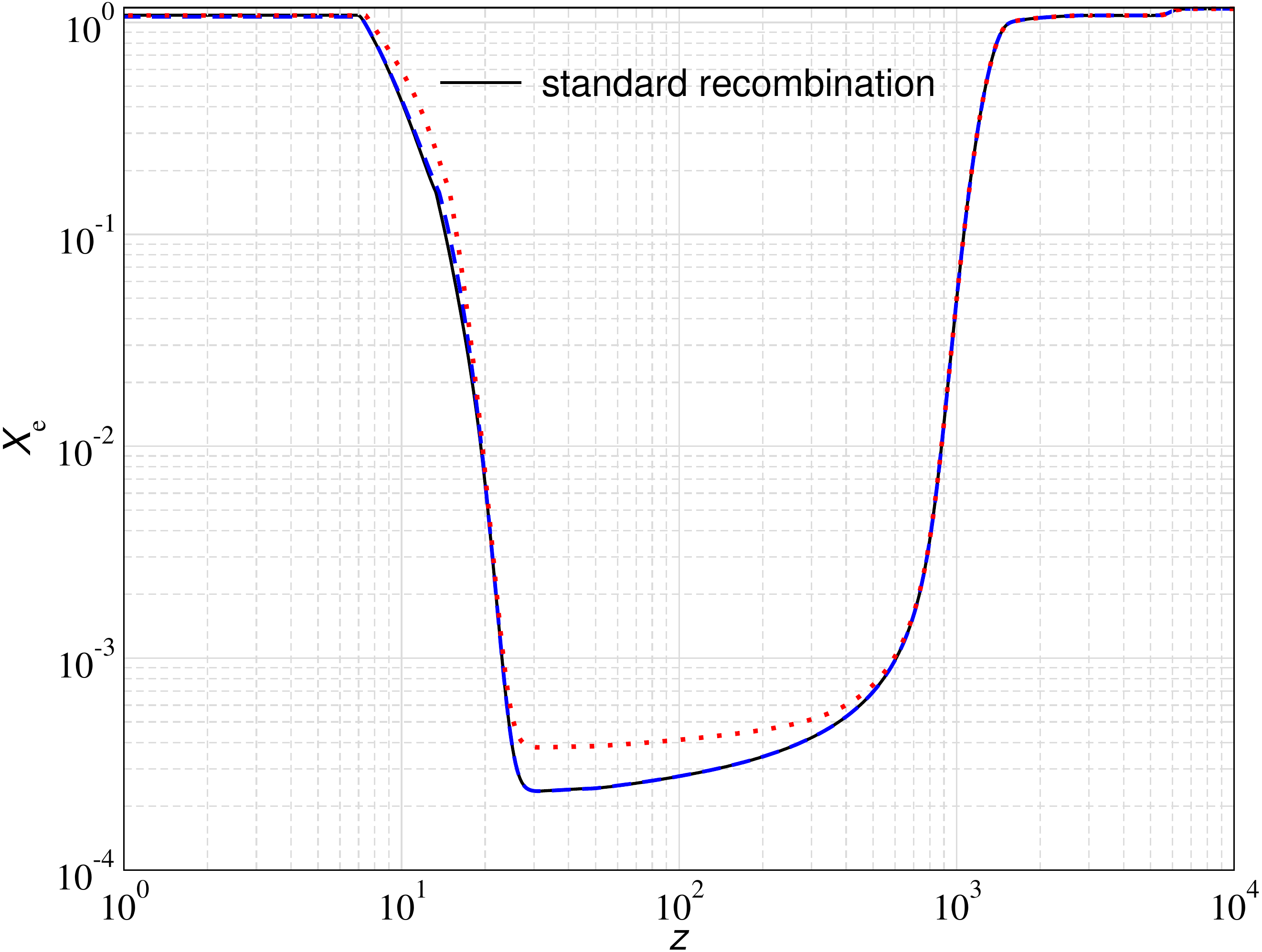}
\hspace{4mm}
\includegraphics[width=\columnwidth]{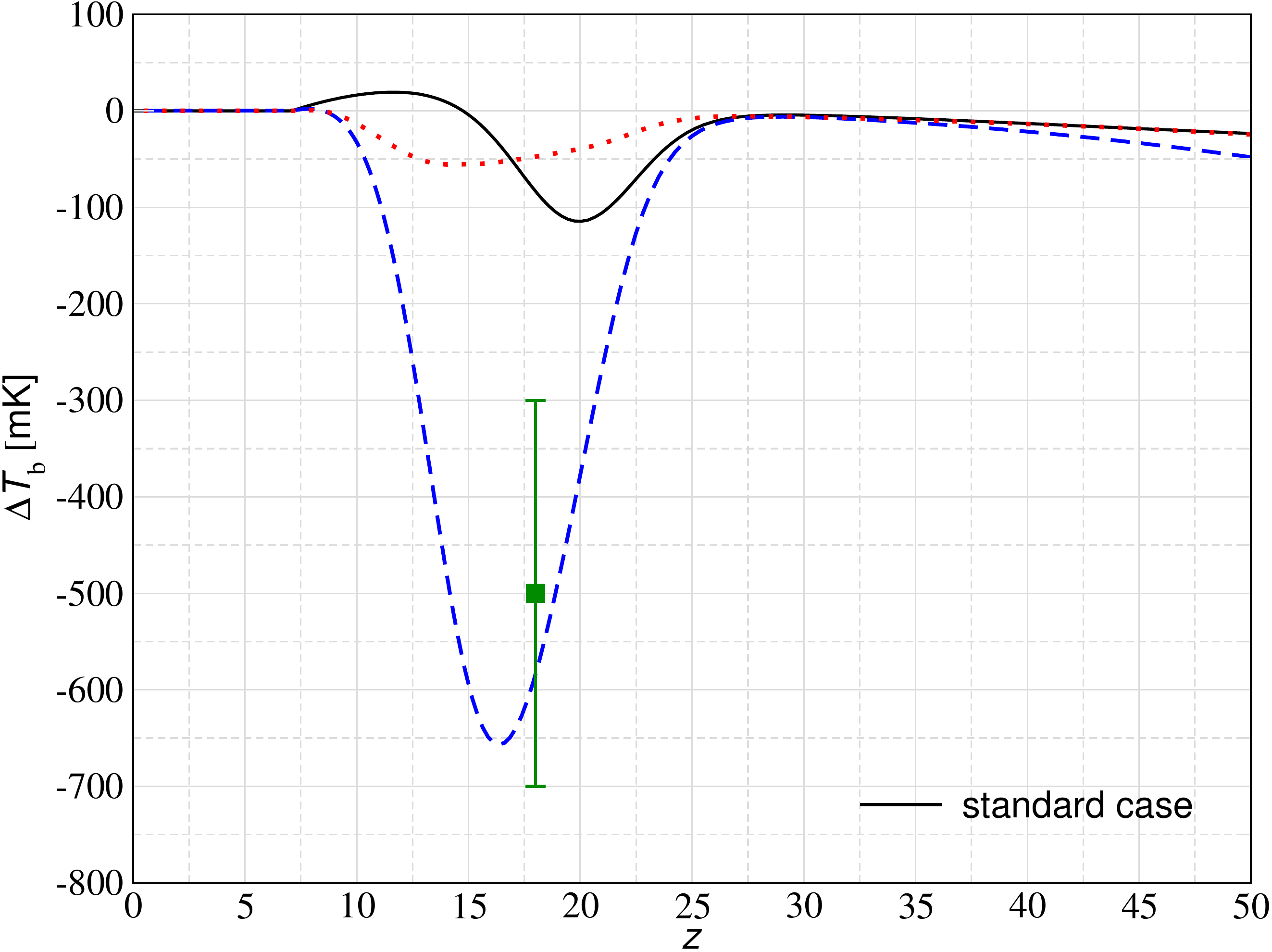}
%\\[10mm]
\caption{Dimensionless intensity (upper left),  equivalent temperature in the Rayleigh-Jeans limit (uuper right), ionization history (bottom left) and the 21 cm distortion (bottom right) for a few parameter combination which fit the data well with free-free-like spectrum with parameter combinations as shown. We have compared with a synchrotron-like solution for reference.}
\label{fig:soln_fit_ff}
\end{figure*}
%--------------------------------------------------

%------------------------------------------------------------
\begin{figure*}
\centering 
\includegraphics[width=\columnwidth]{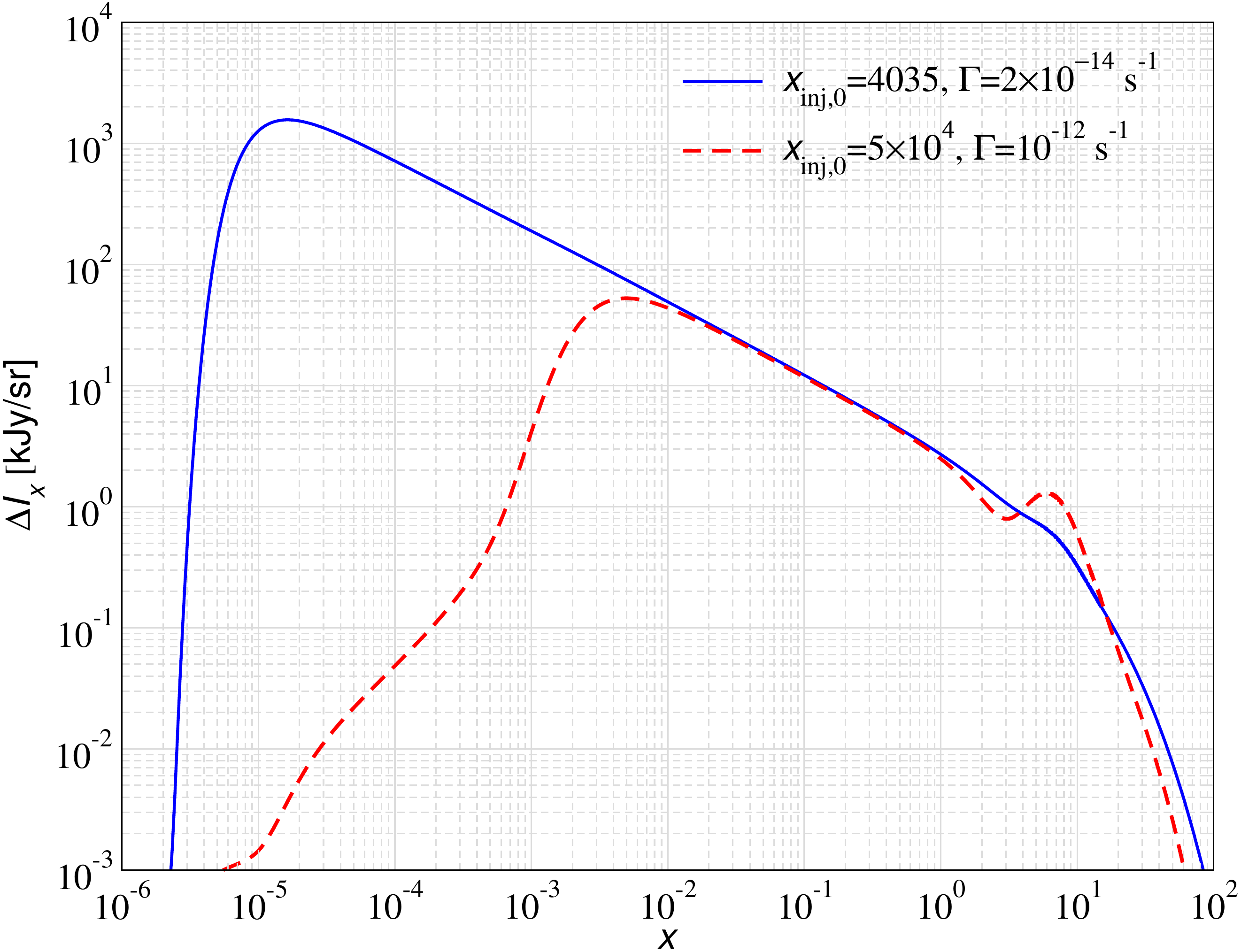}
\hspace{4mm}
\includegraphics[width=\columnwidth]{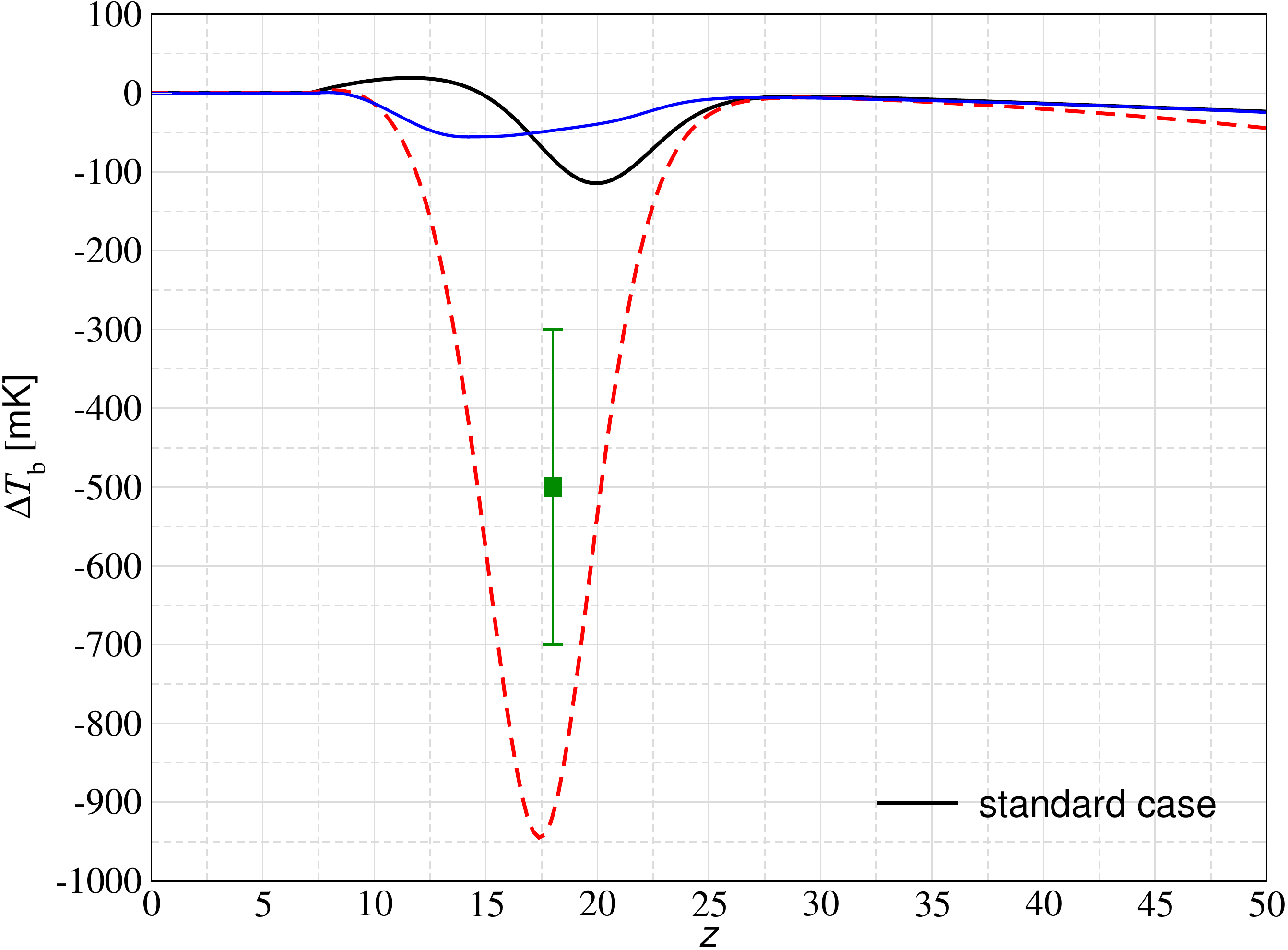}
\\[10mm]
\caption{Dimensionless intensity (left panel) and the 21 cm distortion signature (right panel) for a short (w.r.t recombination epoch) lifetime case, $\Gamma=10^{-12}$~s$^{-1}$. We have compared with a synchrotron solution with both curves having $\Drr=10^{-5}$. We show the EDGES measurement in green. }
\label{fig:short_lifetime_case}
\end{figure*}
%--------------------------------------------------

In this section, we first identify the allowed parameter space for our toy model. We would like to emphasize that deriving parameter constraints is not the main goal of this paper, but this understanding allows us to motivate parameter choices for a few examples.

As we have already explained, we include CMB spectral distortions, CMB anisotropy, the 21 cm absorption signature and the observed RSB as our cosmological data in order to constrain the model parameters. Our model has three main parameters as described in the previous section. We choose a few illustrative values of $\Drr$ and perform a two parameter search, which is again motivated by our intention to explain the basic physics. We remind the reader that $\Drr$ is constrained to be $\lesssim 10^{-5}$ only within the {\it COBE/FIRAS} band (60-600~GHz). For injections outside of this band and/or in the post-recombination era, this constraint does not apply and we do not run into CMB spectral distortions constraints immediately.  

We show the contours of the allowed parameter space for synchrotron-type injections in Fig.~\ref{fig:sync_constraint}. To begin with, we require that we have a fit to the RSB data over the frequency range of $10\,{\rm MHz}-10\,{\rm GHz}$ in order to cover the observed data points. Since 10 GHz is the highest frequency data point, this presents a restriction on the injection frequency today, requiring $x_{\rm inj,0}\gtrsim 0.2$. Therefore, we have chosen a minimal value of $x_{\rm inj,0}=1$. 
%\changeB{We note that the RSB data is noisy at $\sim 10$ GHz, therefore, it is difficult to introduce a precise criterion, although we deem our approach as conservative}. 
We also ignore stimulated decay and non-linear scattering terms in this work, which can become important for lower frequency photons \citep{Largeenergy2022, Bolliet2020PI}. This might affect the details for cases with $x_{\rm inj,0}\lesssim 1$ but again should not change the main conclusions.

We see from the figure that only post-recombination scenarios ($\Gamma\lesssim 10^{-13}$s$^{-1}$) are allowed by the data. We first discuss this region of parameter space and will return to pre-recombination cases later. In a post-recombination universe, energetic photons with energy above the ionization threshold of hydrogen will ionize the neutral atoms. For a synchrotron or free-free type spectrum, most of the energy is in the high energy tail. This implies that for our choices of $\Drr$, the photons will ionize the universe completely as the baryons will be outnumbered by these energetic photons. Therefore, we require that $x_{\rm inj,0}\lesssim 5.8\times 10^4$ ($\equiv$13.6 eV) to avoid strong CMB anisotropy constraints. Reducing $\Drr$ allows us to avoid this constraint, but it then becomes impossible to match the observed amplitude of the RSB. In reality, we find that $x_{\rm inj,0}$ has to be restricted to $\lesssim 2\times 10^4$, otherwise significant direct ionizations will occur. This is due to our choice of spectrum, which has a smooth exponential tail rather than a sharp cutoff. For smaller $x_{\rm inj,0}$, although we avoid direct ionizations, we still see deviations from the standard ionization history (bottom left panel of Fig.~\ref{fig:soln_fit}). This is because the low-frequency photons are absorbed by the residual electrons which leads to heating. This heating then increases in matter temperature, resulting in indirect ionizations, as well as less recombination. As the name "soft" suggests the low frequency photons carry a small fraction of total photon spectrum energy. Therefore, even for $\Drr$ of the order of $10^{-4}-10^{-5}$, we do not violate the cosmological constraints.

In order to fit the RSB data over the broad frequency range, it is necessary to have the correct slope as well as match the normalization. For a synchrotron-like spectrum, the injected soft photons naturally provide a good match to the data (Top right of Fig. \ref{fig:soln_fit}). For a given $\Drr$, there is degeneracy between $x_{\rm inj,0}$ and $\Gamma$ which sets the normalization of the spectrum. This leads to the isolated bands for the allowed regions in Fig. \ref{fig:sync_constraint}. Instead, if we decrease $\Drr$, the change to the normalization can be compensated by the reduction of $x_{\rm inj,0}$. This leads to a shifting of the bands leftwards for decreasing $\Drr$. For lifetimes longer than the age of the universe, only a small fraction of dark matter decays\footnote{Usually $f_{\rm dm}\simeq 10^{-8}-10^{-7}$ for the considered cases.} which results in steepening of the bands with $\Gamma\lesssim 10^{-17}$s$^{-1}$.

In Fig.~\ref{fig:soln_fit}, we show a few examples of synchrotron-type spectrum injections which are consistent with the data. The drop in the intensity at low frequencies (Top left panel) is due to the absorption of soft photons by the ambient medium. The efficiency of this process depends on the electron number density as well as the redshift. At shorter lifetimes, the universe is denser which makes the absorption process more effective \citep{Chluba2015GreensII}. For $x_{\rm inj,0}\simeq 10^4$, one can start to see the effect of direct ionization with a steep rise in the free electron fraction (blue dashed line in the bottom left panel). However, this change is most notable at lower redshifts and is therefore still consistent with CMB anisotropy data. Without heating considerations, the increased radio background should have led to a larger absorption signal compared to the standard scenario, as was pointed out by \cite{FH2018}. Interestingly, we see that with heating included the absorption depth is actually reduced, as is shown in the bottom right panel of Fig.~\ref{fig:soln_fit}. We discuss the interplay of these two opposing effects in more detail in the next section.  

In Fig. \ref{fig:soln_fit_ff}, we show a case of free-free-type injection and compare with the synchrotron case. As expected, this solution is not a very good fit to RSB data even with the inclusion of an irreducible radio background due to a mismatch in the slope of the spectrum and a lack of Comptonization at late times. However, it is still not ruled out at the 2-3$\sigma$ level with the current available data. We show one particular best-fit case to compare with the synchrotron solution in Fig.~\ref{fig:soln_fit_ff}. One could likely improve the fit to the data using gaunt factor corrections to free-free type emissions \citep{BRpack}.  Since free-free has a flatter spectrum compared to synchrotron, there are fewer soft photons which intuitively renders the heating effect more moderate. This also explains why there is almost no modification to the standard ionization history. Interestingly, the 21~cm absorption signal is enhanced for this scenario and somewhat exceeds the EDGES dip, but without violating it strongly. This is expected as the reduction of soft photons implies less heating of the surrounding matter, and therefore a deeper absorption trough.

The soft photon heating quickly loses its importance in the pre-recombination universe ($\Gamma\gtrsim 10^{-13}$s$^{-1}$). During this epoch, the low frequency photons are efficiently absorbed as the universe is completely ionized. This leads to a large decrement in the photon spectrum at low frequencies (left panel of Fig. \ref{fig:short_lifetime_case}). The resultant heating manifests itself as a $y$-distortion, which leads to the bump at $x\simeq 10$. Due to the missing photons, it is not possible to obtain a good fit to the RSB data, especially in the band of 40-80 MHz. This decrement in low-frequency photons implies that the heating effect is almost non-existent at $z\simeq 20$. Therefore, the surviving radio background dominates over any residual heating, and makes the absorption signal deep enough to be in tension with EDGES (right panel of Fig.~\ref{fig:short_lifetime_case}). This behaviour rules out the $\Gamma\gtrsim 10^{-13}$s$^{-1}$ parameter space in Fig.~\ref{fig:sync_constraint}. Nevertheless, it is interesting to note that a radio background can survive from the pre-recombination epoch and can still give a large absorption signal (comparable to EDGES detection). However, we do not pursue this aspect further in this work. In the next section, we describe the soft photon heating effect in more detail.

%------------------------------------------------------------------
\section{Importance of soft photon heating}
\label{sec:soft_heating}
%------------------------------------------------------------------
  
%------------------------------------------------------------
\begin{figure*}
\centering 
\includegraphics[width=\columnwidth]{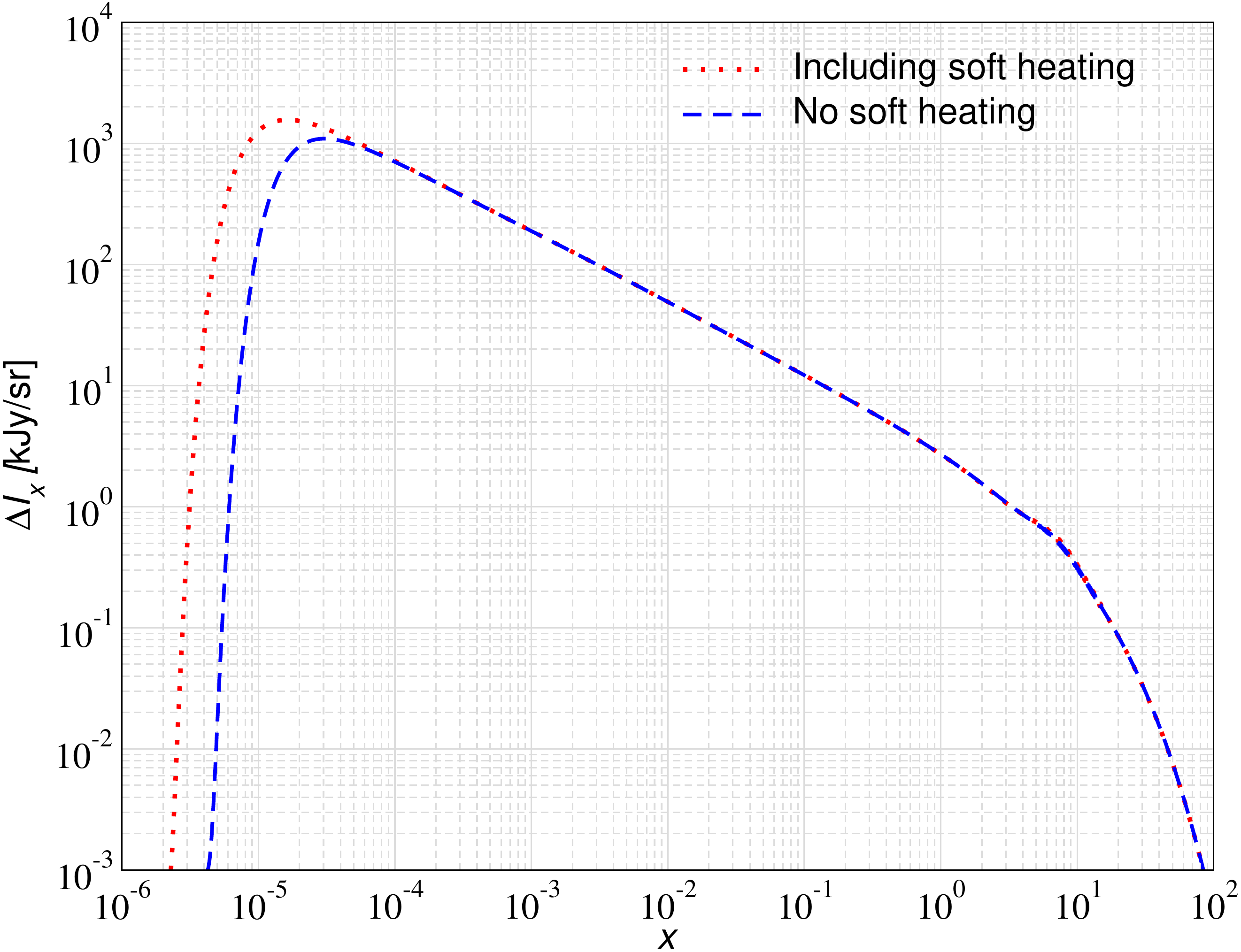}
\hspace{4mm}
\includegraphics[width=\columnwidth]{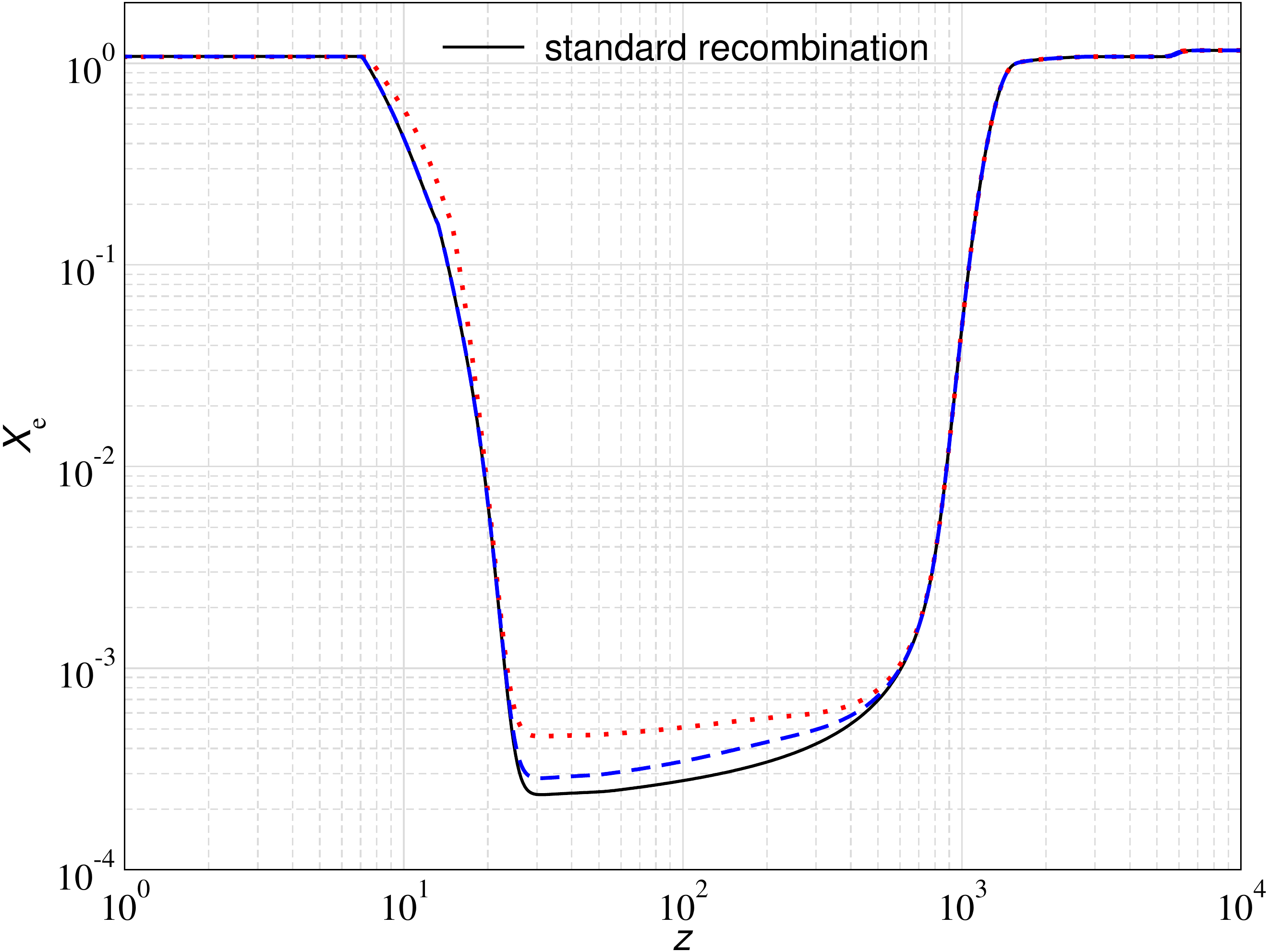}
\\[10mm]
\includegraphics[width=\columnwidth]{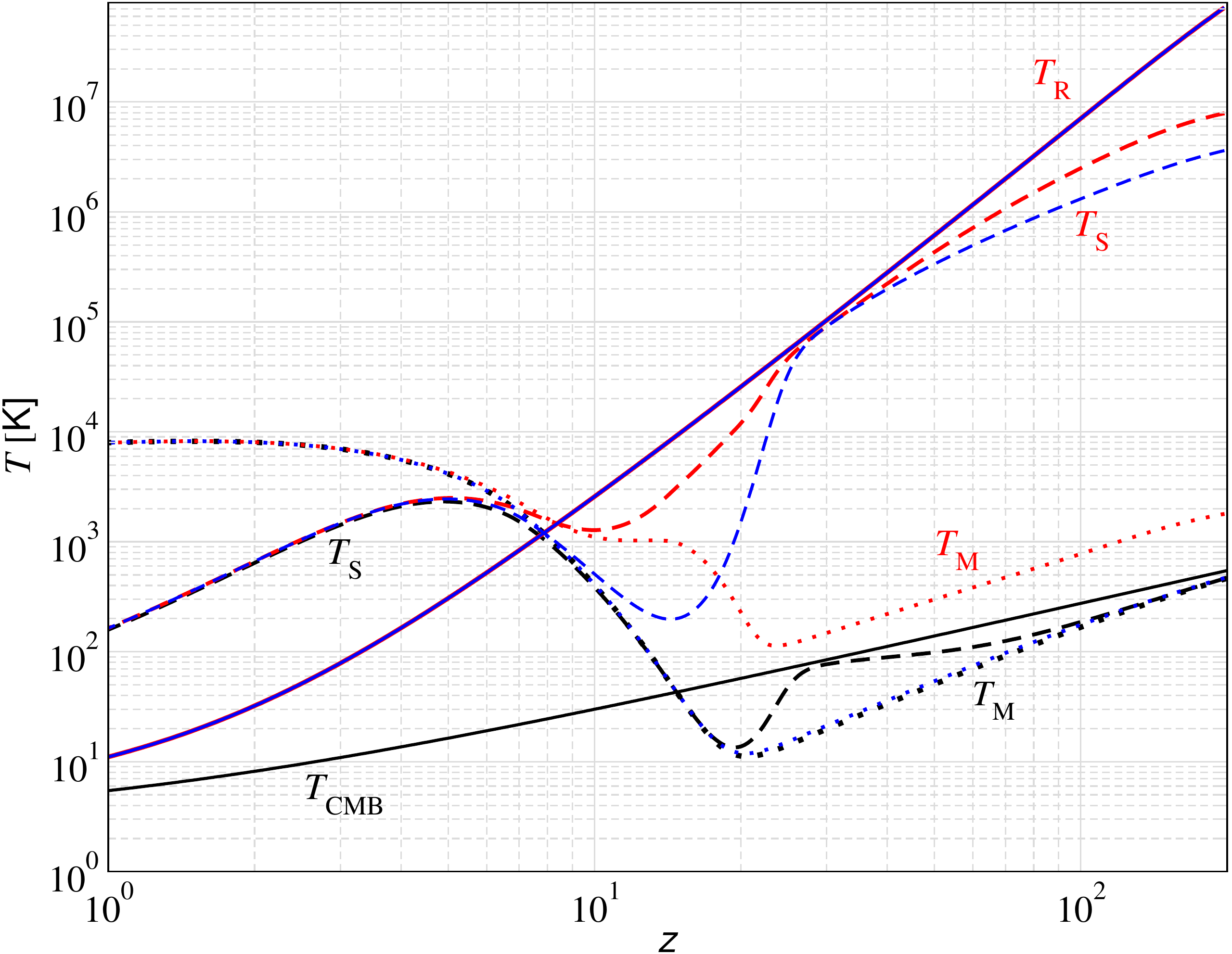}
\hspace{4mm}
\includegraphics[width=\columnwidth]{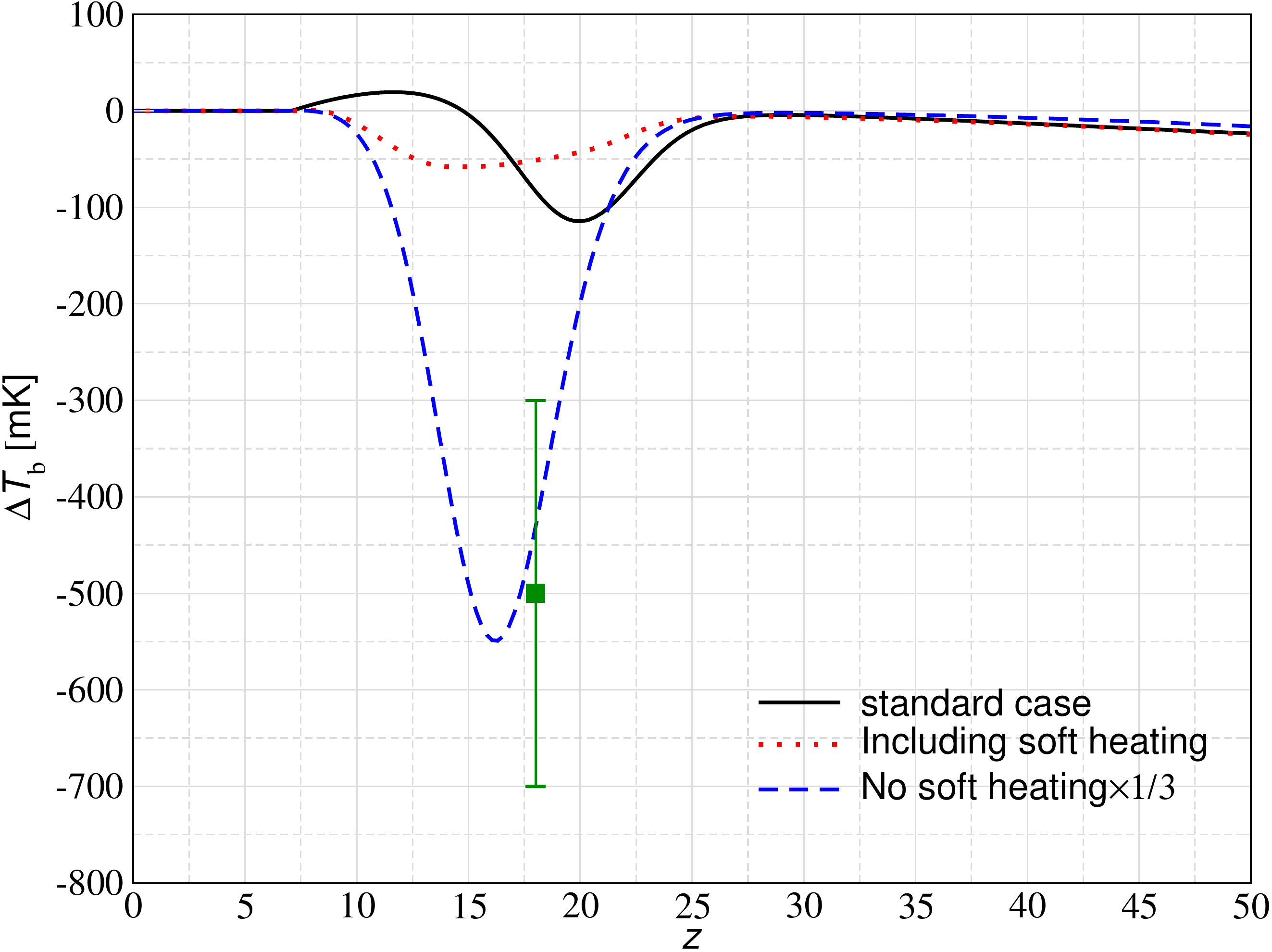}
%\\[10mm]
\caption{Comparison of calculations with (dotted red) and without (dashed blue) the soft photon heating. We choose a synchrotron case with $\Drr=10^{-5}$, $x_{\rm inj,0}=4035$, $\Gamma=2\times 10^{-14}$s$^{-1}$.}
\label{fig:heat_comp}
\end{figure*}
%------------------------------------------------------------

%------------------------------------------------------------
\begin{figure*}
\centering 
\includegraphics[width=\columnwidth]{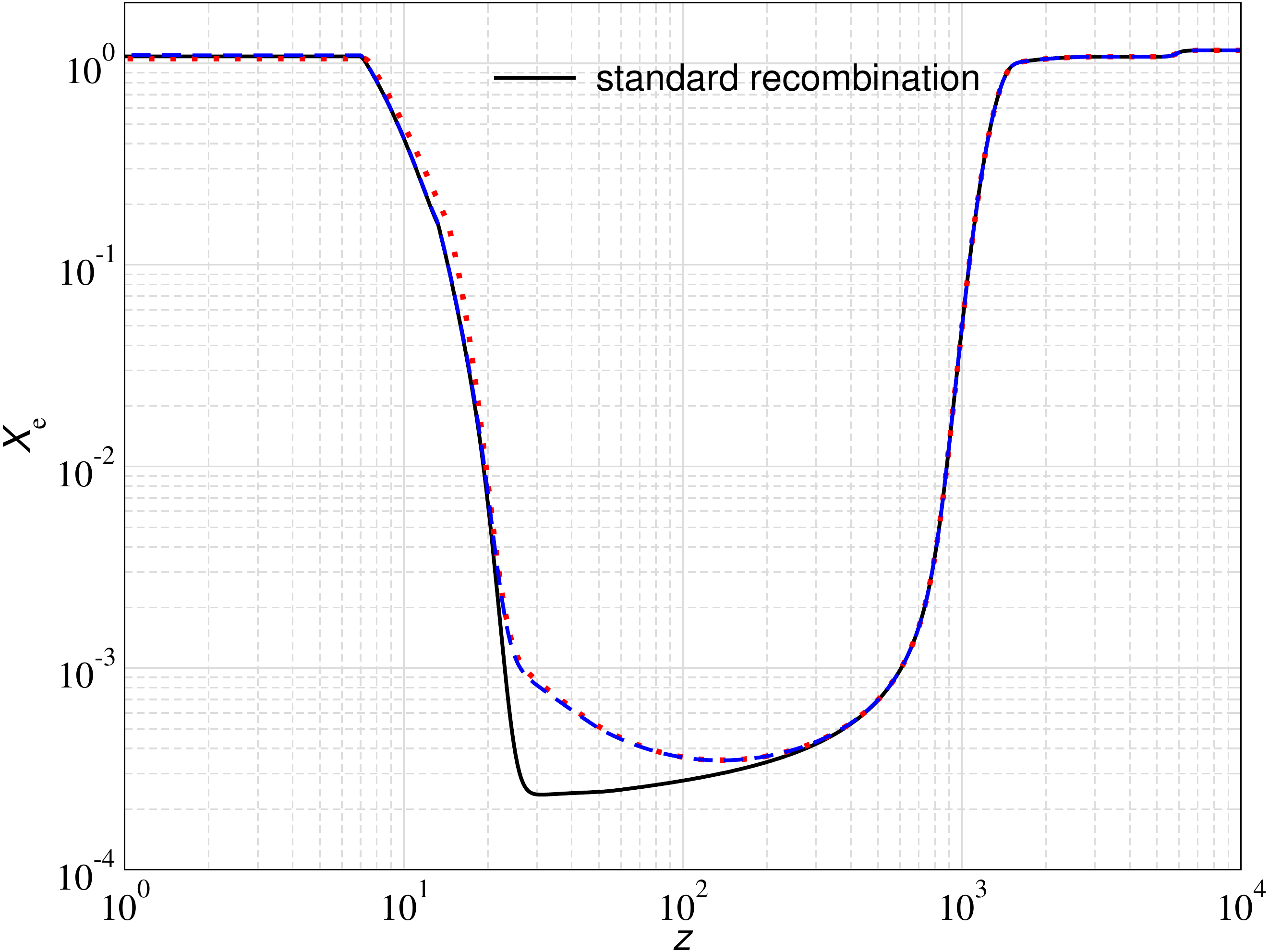}
\hspace{4mm}
\includegraphics[width=\columnwidth]{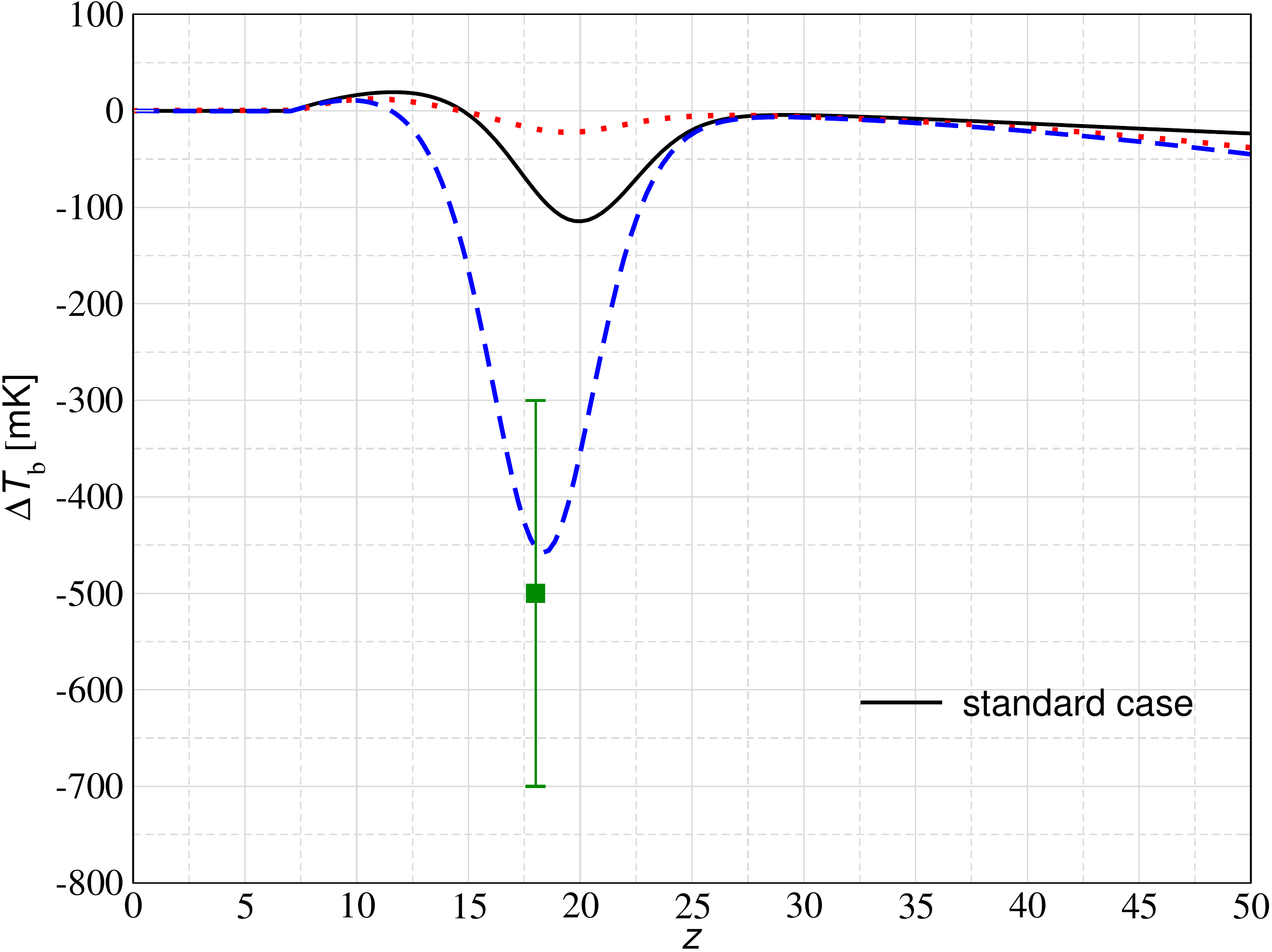}
\\[10mm]
\caption{Ionization history (left panel) and  21 cm distortion signature (right panel) for a long (w.r.t recombination epoch) lifetime case, $\Gamma=10^{-17} $s$^{-1}, x_{\rm inj,0}=10^4$ with both curves having $\Drr=10^{-4}$ with a synchrotron spectrum. The calculations in red dotted include soft photon heating while dashed blue ignores it. }
\label{fig:long_lifetime_case}
\end{figure*}
%--------------------------------------------------

%------------------------------------------------------------
\begin{figure}
\centering 
\includegraphics[width=\columnwidth]{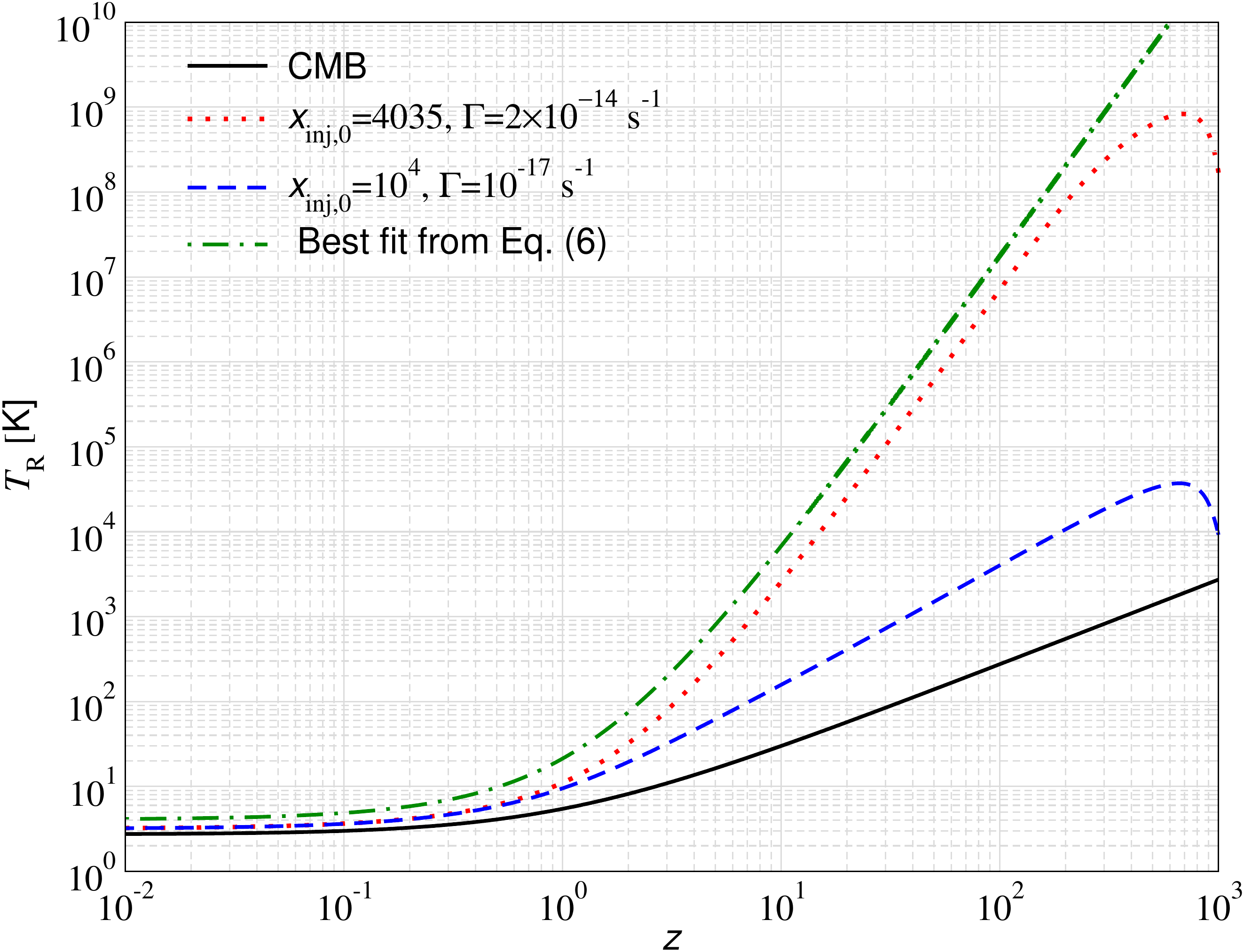}
\hspace{4mm}

\caption{Evolution of $T_{\rm R}$ [K] at 1.4 GHz rest frame frequency for several models that we consider in this work. For comparison we also show the redshifted RSB best-fit given by Eq.~\eqref{eq:RSB-fit}.}
\label{fig:TR_comparison}
\end{figure}
%--------------------------------------------------

In this section, we compare several calculations with and without the soft photon heating effect to showcase its importance for the formation of the global 21~cm signal. 
In practice this means that in {\tt CosmoTherm} we switch the related free-free heating terms off.

The first example is shown in Fig.~\ref{fig:heat_comp}. We consider a case with a long lifetime leading to injection after the recombination epoch, when the soft photon heating effect is most important, as discussed in the previous section. The chosen lifetime is also short enough for the dark matter to have already decayed by $z\simeq 20$ such that the radio background is established well before the 21~cm absorption signal is imprinted. A radio background which is already present at $z\simeq 20$ and simultaneously gives rise to the RSB as seen today, would be in tension with the EDGES detection \citep{FH2018}. Therefore, such a scenario will be ideal to highlight the importance of the new heating effect. We remind the reader that the parameter combinations studied in this and the previous sections are consistent with the cosmological data that we consider in this work.

For the chosen example, the ionization history, is modified noticeably even without the heating (top panel of Fig. \ref{fig:heat_comp}). This is due to some hard photons in the exponential tail which can ionize neutral hydrogen. However, the temperature evolution history is more interesting to look at, as it summarizes the qualitative physics. Soft photon injections naturally lead to a larger radio background compared to the CMB and therefore a larger 21~cm brightness temperature. With the inclusion of soft photon heating, we also have a slightly higher matter temperature at $z\gtrsim 20$ which receives a significant boost during reionization due to availability of free electrons. Indeed, the matter temperature mimics the reionization history with a sharp rise at $z\simeq 10-20$. This boost in the matter temperature reduces the contrast between the brightness and spin temperature. Ignoring the soft photon heating, of course, increases this contrast leading to a deeper absorption signal which is almost three times higher than the EDGES detection ($\simeq -500$ mK) and corroborates the conclusion of \cite{FH2018}.  With the soft photon heating included, we are no longer in tension with EDGES, but the morphology is quite different and not just a scaled version of even the standard case which implies it could be distinguished in future 21~cm observations.   

As a second example, we consider a long lifetime case (compared to the age of universe) in Fig.~\ref{fig:long_lifetime_case}. As we have already argued, for this choice of $x_{\rm inj,0}$, direct ionizations by hard photons lead to a modification of the ionization history. Therefore, switching off soft photon heating does not make a significant difference. Interestingly, even without the heating, this model does not strongly violate both the RSB data and EDGES. This is because in this case the radio background continues to build up even at $z\lesssim 20$. Therefore, a smaller radio background at $z\simeq 20$ can be consistent with EDGES but can also still match the RSB data as observed today. However, the heating effect is important in this case too, resulting in a much smaller absorption signal once accounted for. 

In Fig.~\ref{fig:TR_comparison}, we show the evolution of brightness temperature ($T_{\rm R}$) evaluated at the 1.4 GHz rest frame frequency for the cases that we have considered in Fig.~\ref{fig:heat_comp} and \ref{fig:long_lifetime_case}. For reference, we compare with the RSB best-fit given in Eq.~\eqref{eq:RSB-fit}. We have excluded the irreducible background in the comparison, assuming it is produced independently at $z\lesssim 1$. The figure clearly illustrates that in the short lifetime scenario, the excess brightness is essentially a redshifted version of the RSB, while for the long liftime case, the full RSB is only reached at $z\simeq 0.1$.

In principle, one can additionally heat matter due to the emission of X-rays from structures at $z\lesssim 20$. Typically, however, emission from astrophysical objects has a broad emission spectrum. The X-ray emission is accompanied by UV emission which can ionize neutral hydrogen. Therefore, the production of X-rays will result in early reionization, leading to a tension with CMB data \citep{Planck2018params}. Recently, one such example for accretion onto primordial black holes was studied in \cite{ADC2022}. The soft photon heating avoids this complication, making the 21~cm absorption signal consistent with RSB as well as CMB data. 

%-----------------------------------------------------------------
\section{Conclusion}
\label{sec:conclusion}
%-----------------------------------------------------------------
In this paper, we showcase the importance of soft/radio photon heating to the formation of the 21~cm global signal. This effect is caused by free-free absorption of the injected soft photons and has been neglected in the literature until now. The increased photon brightness at the 21~cm rest frame frequency of 1.4 GHz due to the radio background can be partially compensated by the related heating induced by broadband photon injections. When taken into account, this effect allows us to reconcile the 21~cm absorption detection by EDGES and the observed RSB background, bypassing the limitations raised in \cite{FH2018}. We show that this mechanism can be effective in the post-recombination universe, which helps to bring CMB, 21~cm and RSB data all together within our toy model.

Recently, 21~cm anisotropy measurements \cite{HERA2022} have started to place independent constraints on existing radio backgrounds at $z\simeq 10$. The calculations, however, assume that the radio background can be simply redshifted back, similar to the treatment of \cite{FH2018}. It is clear that this assumption will lead to a higher prediction for the contrast between brightness and spin temperature (Fig~\ref{fig:heat_comp}). In turn, this would boost the 21~cm anisotropy, which these experiments measure in order to place a limit on the amplitude of the ambient radio background. As we have argued here, the soft photon heating could reduce this contrast, which will in turn reduce the predicted anisotropy within a given physical scenario. Therefore, one expects that the constraints will weaken severely with the inclusion of this new effect, unless the additional heating can be excluded by other means.   

We also highlight that the examples considered here (e.g., Fig.~\ref{fig:soln_fit}) all exhibit significant variation in the global 21~cm signal during the cosmic dawn. Given the larger astrophysical uncertainty in the generation of the signal at $z\lesssim 20$, it could be vital to target this range in future observation from the dark side of the moon \citep{Farside2013}. Additional spectral distortion measurements at $10-20\,{\rm GHz}$, e.g. with TMS \citep{Jose2020TMS} could furthermore shed new light on the physics causing the ARCADE excess.

We would also like to add that the calculations performed in this work represent a simplified analysis. For instance, we have assumed that the problem remains linear. However, a more thorough calculation without this assumption may be necessary following \cite{Chluba2020large,Largeenergy2022}.
In the linear problem, we have assumed that only the background CMB photons stimulate the scattering of non-thermal soft photons. However, the intensity of this non-thermal background can be orders of magnitude higher than the CMB photons at least in the low frequency regime. Therefore, one may also have to take into account stimulation by the non-thermal soft photons themselves, which also makes the emission problem non-linear \citep{Brahma2020, Bolliet2020PI, AC2022}. We defer this aspect of the problem to future work.

Lastly, even though we have assumed a relatively simple phenomenological model in our calculations, the qualitative conclusions about the importance of soft photon heating will apply to any model which injects a broad spectrum of radio photons. This might be important for astrophysical objects such as radio galaxies which have recently gathered attention as a significant source of radio emission \citep{RFB2020}. A follow-up to this work in the context of emission from superconducting strings is currently in progress (Cyr et. al., in preparation), indicating that even some early-universe models may provide novels ways of thinking about the creation of CMB spectral distortions and 21~cm signals as simply two sides of the same coin.

\section*{Acknowledgments}

This work was supported by the ERC Consolidator Grant {\it CMBSPEC} (No.~725456).
JC was furthermore supported by the Royal Society as a Royal Society University Research Fellow at the University of Manchester, UK (No.~URF/R/191023).
BC would also like to acknowledge support from an NSERC-PDF.

\vspace{-5mm}
\section{Data availability}
The data underlying in this article are available in this article and can further be made available on request.

{\small
\vspace{-3mm}
\bibliographystyle{mn2e}
\bibliography{Lit}
}
\newpage

%\appendix

%\section{appendix}

\end{document}